\documentclass[fleqn,usenatbib]{mnras}

\usepackage[T1]{fontenc}
\usepackage{ae,aecompl}
 
\usepackage{graphicx}	% Including figure files
\usepackage{amsmath}	% Advanced maths commands

\usepackage{amssymb}	% Extra maths symbols
\usepackage{float}
\usepackage{caption}
\usepackage{subcaption}

\title[Radial profiles of the Virgo Cluster. ]{ Chemical enrichment of the ICM within the Virgo cluster I: radial profiles.}

\author[Gatuzz et al.]{
Efrain Gatuzz$^{1}$\thanks{E-mail: egatuzz@mpe.mpg.de},
J. S. Sanders$^{1}$,
K. Dennerl$^{1}$,
A. Liu$^{1}$,
A. C. Fabian$^{2}$,
C. Pinto$^{3}$,\newauthor
D. Eckert$^{4}$,
H. Russell$^{5}$,
T. Tamura$^{6}$,
S. A. Walker$^{7}$
and J. ZuHone$^{8}$
\\
$^{1}$ Max-Planck-Institut f\"ur extraterrestrische Physik, Gie{\ss}enbachstra{\ss}e 1, 85748 Garching, Germany\\
$^{2}$ Institute of Astronomy, Madingley Road, Cambridge CB3 0HA, UK\\ 
$^{3}$ INAF - IASF Palermo, Via U. La Malfa 153, I-90146 Palermo, Italy \\
$^{4}$ Department of Astronomy, University of Geneva, Ch. d\rq Ecogia 16, CH-1290 Versoix, Switzerland \\
$^{5}$ School of Physics \& Astronomy, University of Nottingham, University Park, Nottingham NG7 2RD, UK\\ 
$^{6}$ Institute of Space and Astronautical Science (ISAS), Japan Aerospace Exploration Agency (JAXA) Kanagawa 252-5210, Japan\\  
$^{7}$ Department of Physics and Astronomy, University of Alabama in Huntsville, Huntsville, AL 35899, USA\\
$^{8}$ Harvard-Smithsonian Center for Astrophysics, 60 Garden Street, Cambridge, MA, 02138, USA
}

\date{Accepted XXX. Received YYY; in original form ZZZ} 
\pubyear{2018} 
\begin{document}
 \label{firstpage}
\pagerange{\pageref{firstpage}--\pageref{lastpage}}
\maketitle 

\begin{abstract}
We present a detailed analysis of the elemental abundances distribution of the Virgo cluster using {\it XMM-Newton} observations. We included in the analysis a new EPIC-pn energy scale calibration which allow us to measure velocities with uncertainties down to $\Delta v \sim 150$ km/s. We investigate the radial distribution of O, Ne, Mg, Si, Ar, S, Ca, Ni and Fe. We found that the best-fit model is close to a single-temperature component for distances $>80$~kpc and the cooler gas is more metal-rich. Discontinuities in temperature are found around $\sim30$~kpc and $\sim90$~kpc, which correspond to the radius of the cold fronts. We modeled elemental X/Fe ratio profiles with a linear combination of SNIa and SNcc models. We found a flat radial distribution of SNIa ratio over the total cluster enrichment, which supports an early ICM enrichment scenario, with most of the metals present being produced prior to clustering.
\end{abstract} 
 


\section{Introduction}\label{sec_in}    
Most of the baryonic mass in clusters of galaxies lies in the form of a tenuous hot plasma known as intracluster medium (ICM). The chemical composition and physical properties of this environment contains valuable information about the origin and distribution of chemical elements during the evolution of the Universe. Fe-peak elements (Cr, Mn, Fe, Ni) originate from type~Ia supernovae (SNIa) while light $\alpha$-elements (O, Ne, Mg) mainly originate from core-collapse Sne (SNcc). Intermediate-mass elements (e.g. Si, S, Ar and Ca) are synthesized by both SNcc and SNIa \citep[e.g.,][and references therein]{nom13}. The ICM is filled with these elements from their host galaxies by ram-pressure stripping, galactic winds and AGN bubbles uplift. Therefore, the ICM metal is sensitive to both the time-scale over which the supernova products are expelled and the number of SNIa and SNcc that contribute to the chemical enrichment. Other important parameters include the initial mass function (IMF) of the stars that explode as SNcc, the initial metallicity of the progenitors and the SNIa explosion mechanism \citep[see][for a review]{wer08}. 

X-ray spectroscopy of the ICM constitutes a powerful tool to study the abundance distribution of these elements \citep[e.g.,][for a recent review]{mer18}. A central Fe abundance excess was first reported in the Centaurus cluster \citep{all94,fuk94}. Similar trends have been found in many cool-core clusters \citep[e.g.][]{deg01,chu03,pan15,mer17,liu19,liu20}. Recent Perseus cluster observations by {\it Hitomi} indicated that the abundance ratios near the cluster core are fully consistent with solar \citep{hit18,sim19}, thus pointing out the importance of the contribution from both SNIa (with sub-Chandrasekhar mass) and near SNcc to the chemical enrichment of the ICM.

Here, we present an analysis of the chemical enrichment of ICM within the Virgo galaxy cluster. Located at $z=0.00436\pm 0.000023$ \citep{all14}, and being the second-brightest extended extragalactic soft X-ray source, constitutes an excellent laboratory to study the ICM chemical enrichment. The active AGN in its core displays a central jet as well as extended radio bubbles \citep{owe00}, creating through feedback large cavities and shocks in the ICM \citep{for07}. {\it XMM-Newton} observations of the system shows that the X-ray emitting gas form arms with similar abundance ratios of elements such as O, Si and S in and outside such arms, indicating that the metals must have been transported after the last major epoch of star formation \citep{sim07,sim08}. Cold fronts have been identified within the system, with sloshing of gas as the most viable explanation for their presence \citep{sim10,urb11}. Recent analysis of the velocity structure within the Virgo cluster shows signatures for both AGN outflows and gas sloshing \citep{gat22a}.

This paper is organized as follow. In Section~\ref{sec_dat} we describe the data reduction process.  In Section \ref{sec_fits} we explain the fitting procedure. A discussion of the results is shown in Section~\ref{sec_dis} while the conclusions and summary are included in Section~\ref{sec_con}. Throughout this paper we assumed a $\Lambda$CDM cosmology with $\Omega_m = 0.3$, $\Omega_\Lambda = 0.7$, and $H_{0} = 70 \textrm{ km s}^{-1}\ \textrm{Mpc}^{-1} $.

\section{Data reduction}\label{sec_dat}
The {\it XMM-Newton} European Photon Imaging Camera \citep[EPIC,][]{str01} spectra were reduced with the Science Analysis System (SAS\footnote{\url{https://www.cosmos.esa.int/web/xmm-newton/sas}}, version 19.1.0). The observations are the same as used in \citet{gat22a} and we followed the same data reduction process. First, we processed each observation with the {\tt epchain} SAS tool. We used only single-pixel events (PATTERN==0) while bad time intervals were filtered from flares applying a 1.0 cts/s rate threshold. In order to avoid bad pixels or regions close to CCD edges we filtered the data using FLAG==0. 

Following the work done in \citet[][c]{san20,gat22a,gat22b}, we used updated calibration files which allows to obtain velocity measurements down to 150 km/s at Fe-K by using the background X-ray lines identified in the spectra of the detector as references for the absolute energy scale. Identification of point sources was performed using the SAS task {\tt edetect\_chain}, with a likelihood parameter {\tt det\_ml} $> 10$. The point sources were excluded from the subsequent analysis, including the AGN in the cluster core (i.e. a central circular region with a diameter  $D=58$ \arcsec).

In order to study the distribution of chemical elements in the ICM we analyzed non-overlapping circular regions. The thickness of these rings increases as the square root with distance from the M87 center. Figure~\ref{fig_regions} shows the extracted regions. 
    
\section{Spectral fitting}\label{sec_fits}   

Previous analysis of the Virgo cluster have shown the presence of a multi-temperature component near the cluster center \citep{bel01,mol02,mat02,sim07}. In order to avoid switching between single and multi-temperature models for the ICM X-ray emission in some arbitrary way, we decided to use the {\tt lognorm} model \citep{fra13,gat22b}. The {\tt lognorm} model includes a log-normal temperature distribution and takes as input a central temperature ($T$), width of the temperature distribution in log space ($\sigma$), metallicities, redshift and normalization.  The relative normalizations of each component are scaled so that the total is the overall normalization. We included a {\tt tbabs} component \citep{wil00} to account for the Galactic absorption. The free parameters in the model are the temperature, log($\sigma$), elemental abundances (O, Ne, Si, Ar, S, Ca, Fe, Ni) and normalization. Although Al and Mg abundances are free parameters as well, their abundances may not be reliable due to the presence of an strong Al K$\alpha$ instrumental line $\sim 1.5$ keV. We fixed the column density to $1.27\times 10^{20}$ cm$^{-2}$ \citep{wil13}. Different models were also tested, including $N({\rm H})$ as free parameter, which are described in Appendix~\ref{sec_multi_fit}.

\begin{figure}    
\centering
\includegraphics[width=0.40\textwidth]{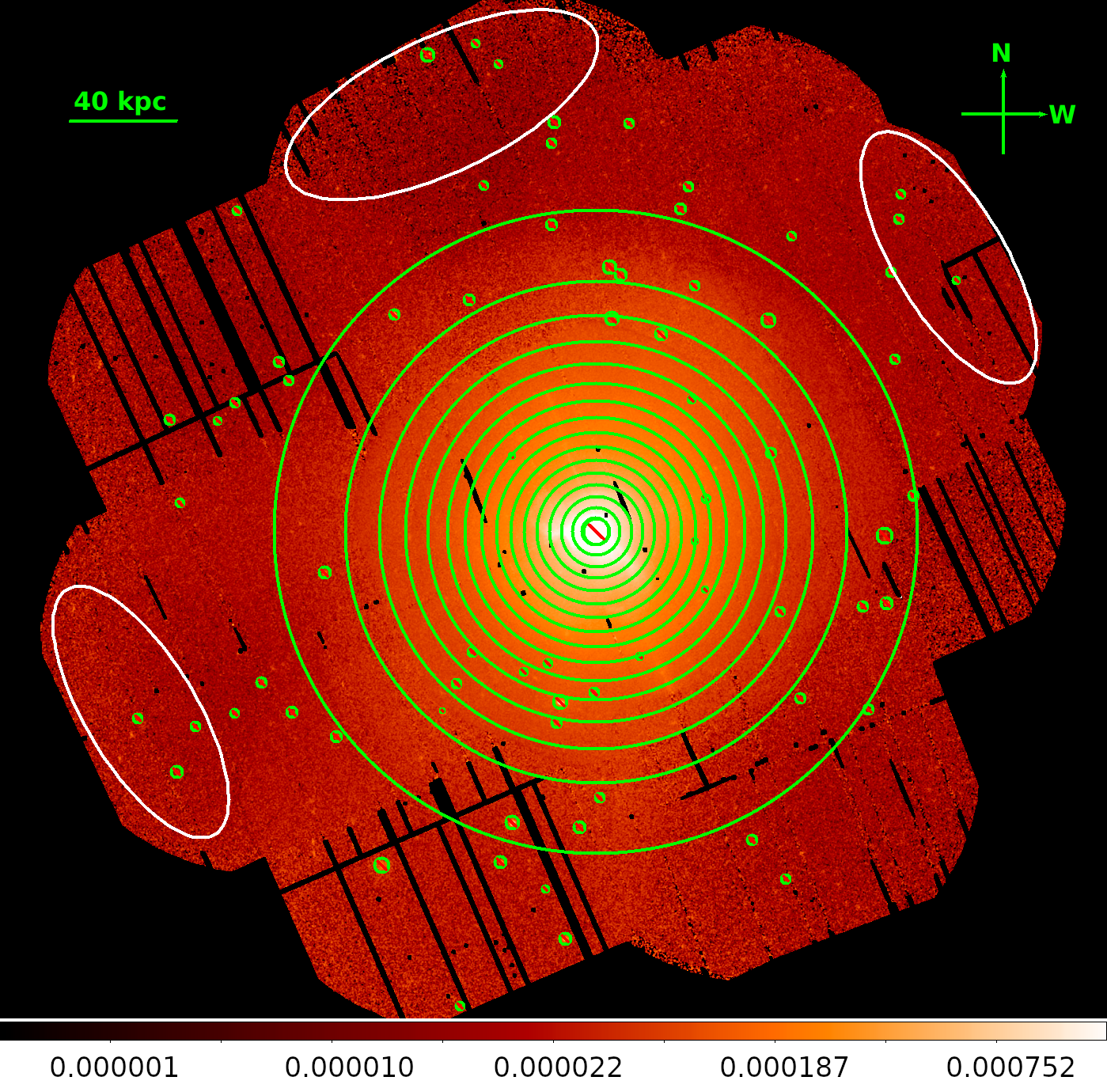} 
\caption{Virgo cluster extracted regions. Black circles correspond to point sources which were excluded from the  analysis, including the AGN in the cluster core. The white regions were used to model the astrophysical background.} \label{fig_regions} 
\end{figure}

As background components we have included Cu-$K\alpha$, Cu-$K\beta$, Ni-$K\alpha$, Zn-$K\alpha$ and Al-$\alpha$ instrumental emission lines, and a powerlaw component with its photon index fixed at 0.136 \citep[the average value obtained from the archival observations analyzed in ][]{san20}. We also take into account the astrophysical background with a power-law with $\Gamma=1.45$ that accounts for the unresolved population of point sources, one absorbed thermal plasma model ({\tt apec}) for the Galactic halo (GH) emission and an unabsorbed thermal plasma model for the Local Hot Bubble (LHB) emission \citep[e.g.][]{yos09}. In order to estimate the temperatures of these components we model the spectrum from 3 elliptical regions located in the outskirt of the cluster, $\sim 30$~arcmin from the cluster center (see Figure~\ref{fig_regions}). Table~\ref{tab_bkg} list the best-fit parameters obtained for the CXB, GH and LHB components. Henceforth, the temperatures for these background components were fixed to the best-fit values of $T_{e}=0.48$~keV and $T_{e}=0.065$~keV. We note that our extraction regions are closer to the cluster core and therefore may be contaminated by an elevated soft foreground component, especially affecting abundance elements of lighter elements such as O.

\begin{table}
\caption{Background model best-fit parameters. }\label{tab_bkg}
\begin{center}
\begin{tabular}{ccccccc}
\hline
\hline
  & $\Gamma$ & & kT (keV)     \\
\hline  
CXB & 1.45  & GH & $ 0.48 \pm 0.02 $   \\
 &  &  LHB & $0.065\pm 0.008$     \\
\hline 
\end{tabular}
\end{center}
  
\end{table}

We combine the spectra from different observations together for each spatial region analyzed and rebinned the spectra to have at least 1 count per channel. Then, we load the data twice in order to fit separately, but simultaneously, the soft 0.5-4.0 keV and the harder 4.0-10 keV energy bands. Given that for lower energies the new EPIC-pn energy calibration scale cannot be applied, the redshift is a free parameter only for the 4.0-10 keV energy band while the inclusion of lower energy band data leads to a better constrain for the metallicities and temperatures. It is important to note that the previous work done in \citet{gat22a} did not included the soft band (i.e. $<4$ keV). We analyze the spectra with the {\it xspec} spectral fitting package (version 12.11.1\footnote{\url{https://heasarc.gsfc.nasa.gov/xanadu/xspec/}}). We assumed {\tt cash} statistics \citep{cas79}. Errors are quoted at 1$\sigma$ confidence level unless otherwise stated. Finally, abundances are given relative to \citet{lod09}.
 
\begin{figure}    
\centering
\includegraphics[width=0.48\textwidth]{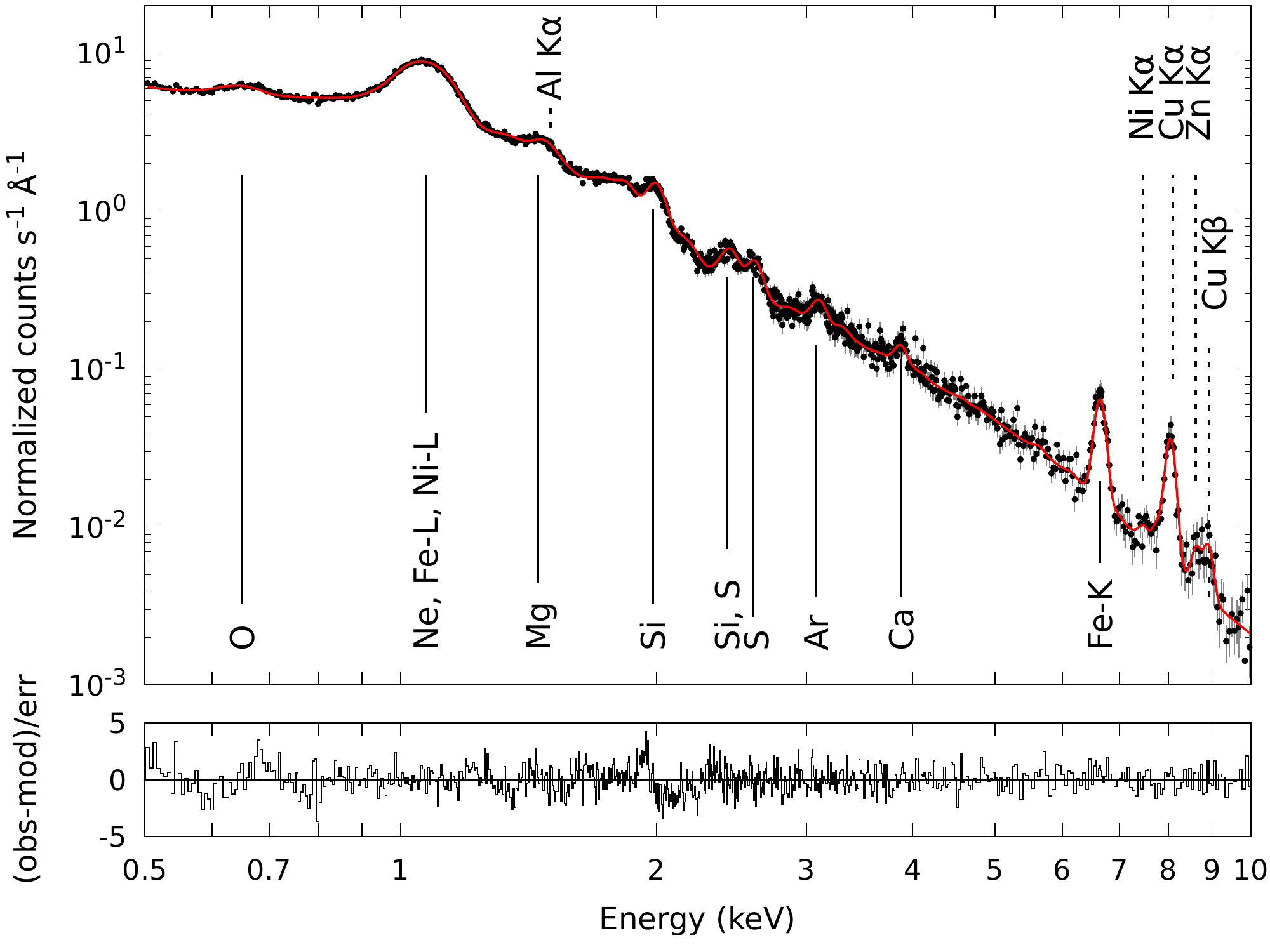} 
\caption{Example spectrum and residuals from the radial analysis. The spectrum has been rebinned for illustrative purposes. The line contribution from ICM emission (vertical solid lines) and from instrumental background (vertical dashed lines) are indicated. The lower panel shows the residuals to the fit.} \label{fig_spectra_example} 
\end{figure}

Figure~\ref{fig_spectra_example} shows an example spectrum and residuals from this analysis. The best-fitting {\tt lognorm} model is indicated as a solid red line. The line contributions of O, Ne, Mg, Si, S, Ar, Ca, Fe and Ni are labelled on the plot. Vertical dashed lines indicate the instrumental Ni K$\alpha$, Cu K$\alpha$,$\beta$, Zn K$\alpha$, Al K$\alpha$  background lines.

\section{Results and discussion}\label{sec_dis}

\subsection{Temperature profile}\label{spec_maps}  
\begin{figure}    
\centering
\includegraphics[width=0.45\textwidth]{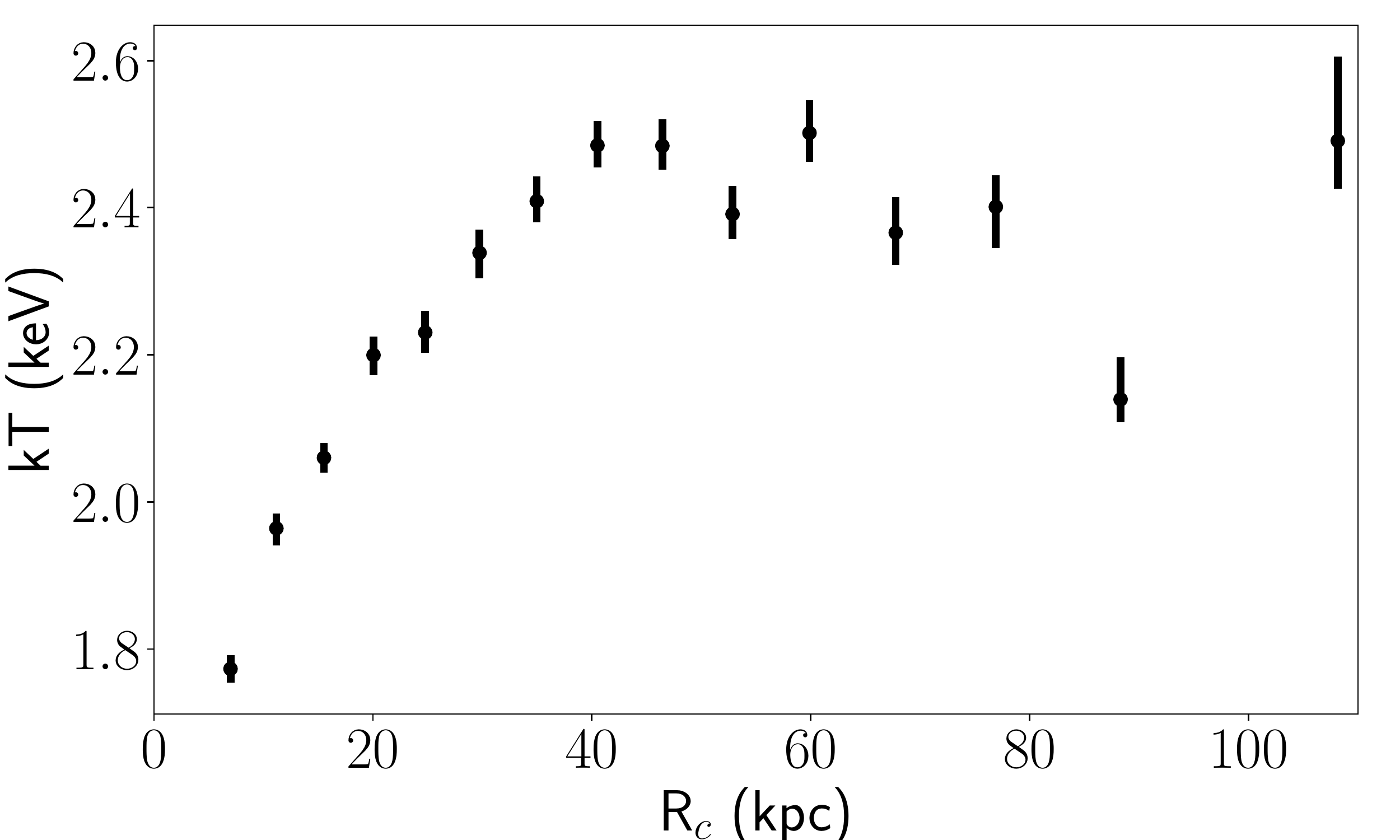}\\
\includegraphics[width=0.45\textwidth]{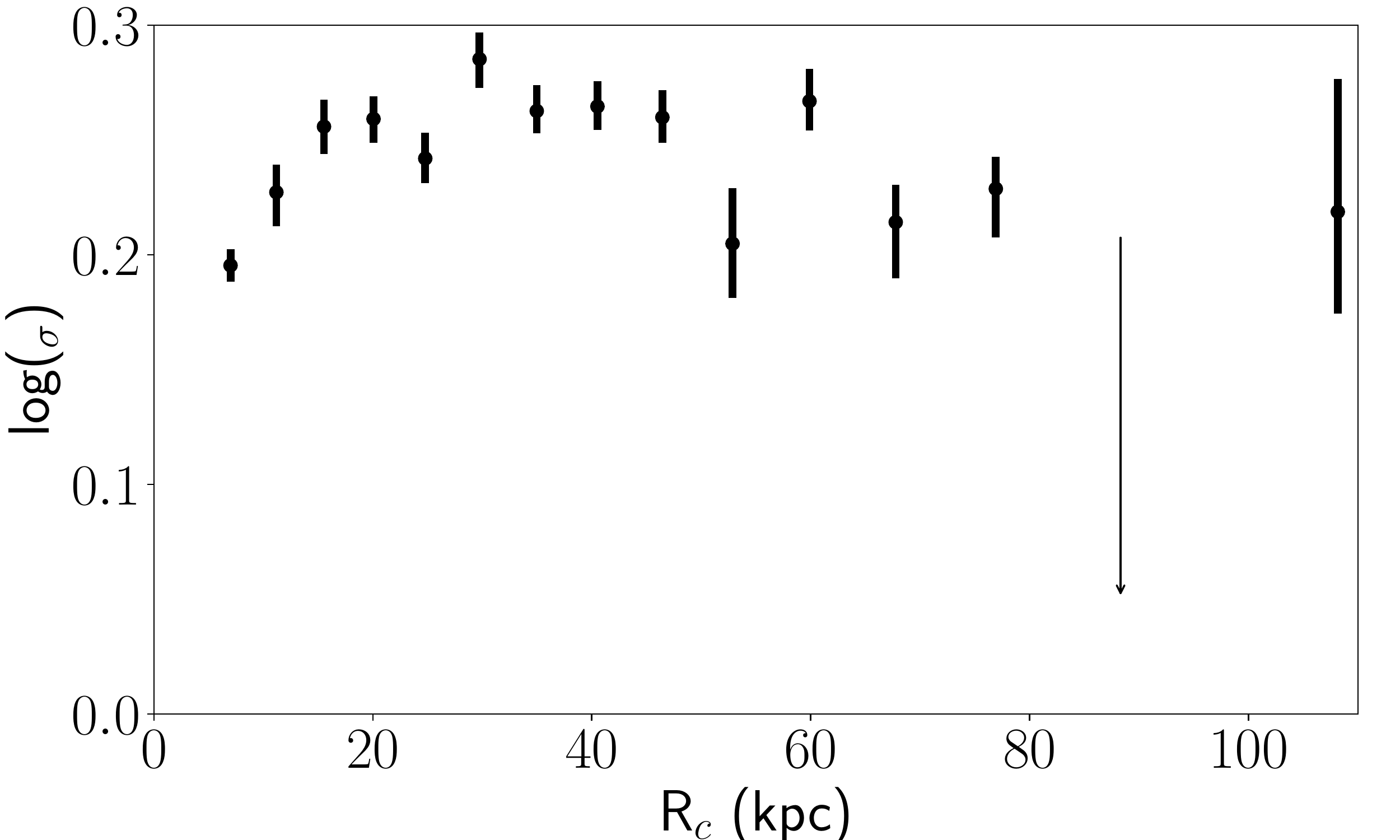} 
\caption{Temperature (top panel) and $\log(\sigma)$ (bottom panel) profiles obtained from the best fit results.} \label{fig_kt_sigma} 
\end{figure}

 As indicated by previous works, a multi-temperature structure with a continuous distribution is more realistic than a single temperature model due to the mixing of gas at several radii during the rise of the radio lobes \citep{chu01,bru02,sim08}. Figure~\ref{fig_kt_sigma} shows the temperatures and $\log(\sigma)$ obtained from the best fit per region. The temperatures steady increase up to $\sim2.45$~keV (at $\sim40$~kpc). A slight increase in temperature may be identified around $\sim$14~kpc, near the shock front. In general, the temperatures are lower than those obtained by \citet[][see Figure~6]{gat22a}. This is expected given that the current analysis includes the soft-energy band. We noted that for distances $\sim 90$~kpc the best-fit model is close to a single-temperature component. There are hints for discontinuities in temperature around $\sim30$~kpc and $\sim90$~kpc, which corresponds to the radius of the inner and outer cold fronts \citep{sim07,sim10}.
 
\subsection{Velocity profile}\label{spec_maps}  
\begin{figure}    
\centering 
\includegraphics[width=0.47\textwidth]{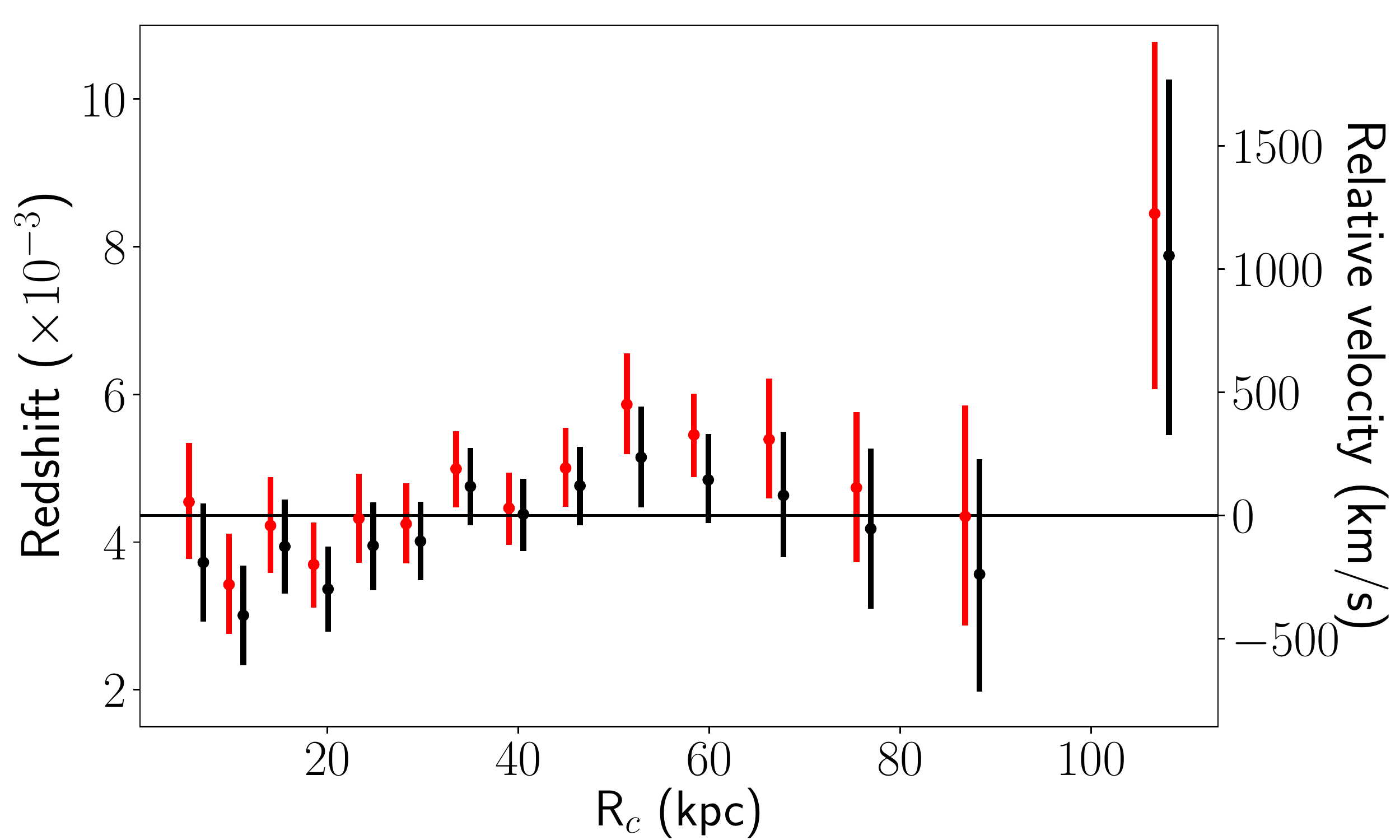}  
\caption{Velocities obtained for each region (black points) and those obtained by \citet[][red points]{gat22a}. The M87 redshift is indicated with an horizontal line. }\label{fig_vel} 
\end{figure}

Figure~\ref{fig_vel} shows the velocities obtained for each region. We have obtained accurate velocity measurements down to $\Delta v\sim 155$ km/s (for ring 7). The largest blueshift/redshift with respect to M87 correspond to $-453_{-207}^{+199}$~km/s and $751_{-837}^{+846}$~km/s for rings 2 and 15, respectively. The blueshifted material tend to have lower temperatures ($<2.3$~keV) and correspond to the increasing part of the temperature profile (see Figure~\ref{fig_kt_sigma}). However, this correlation may be coincidental. The material with larger velocity, on the other hand, has higher temperatures ($>2.3$ keV). Figure~\ref{fig_vel} includes the velocities obtained by \citet{gat22a}. We have found a good agreement even though the current analysis includes the soft-energy band in the spectral fitting, allowing for a better constraining of the temperature which affects the shape of the Fe complex.

\subsection{Abundance profiles}\label{circle_rings} 
\begin{figure*}    
\centering
\includegraphics[width=0.33\textwidth]{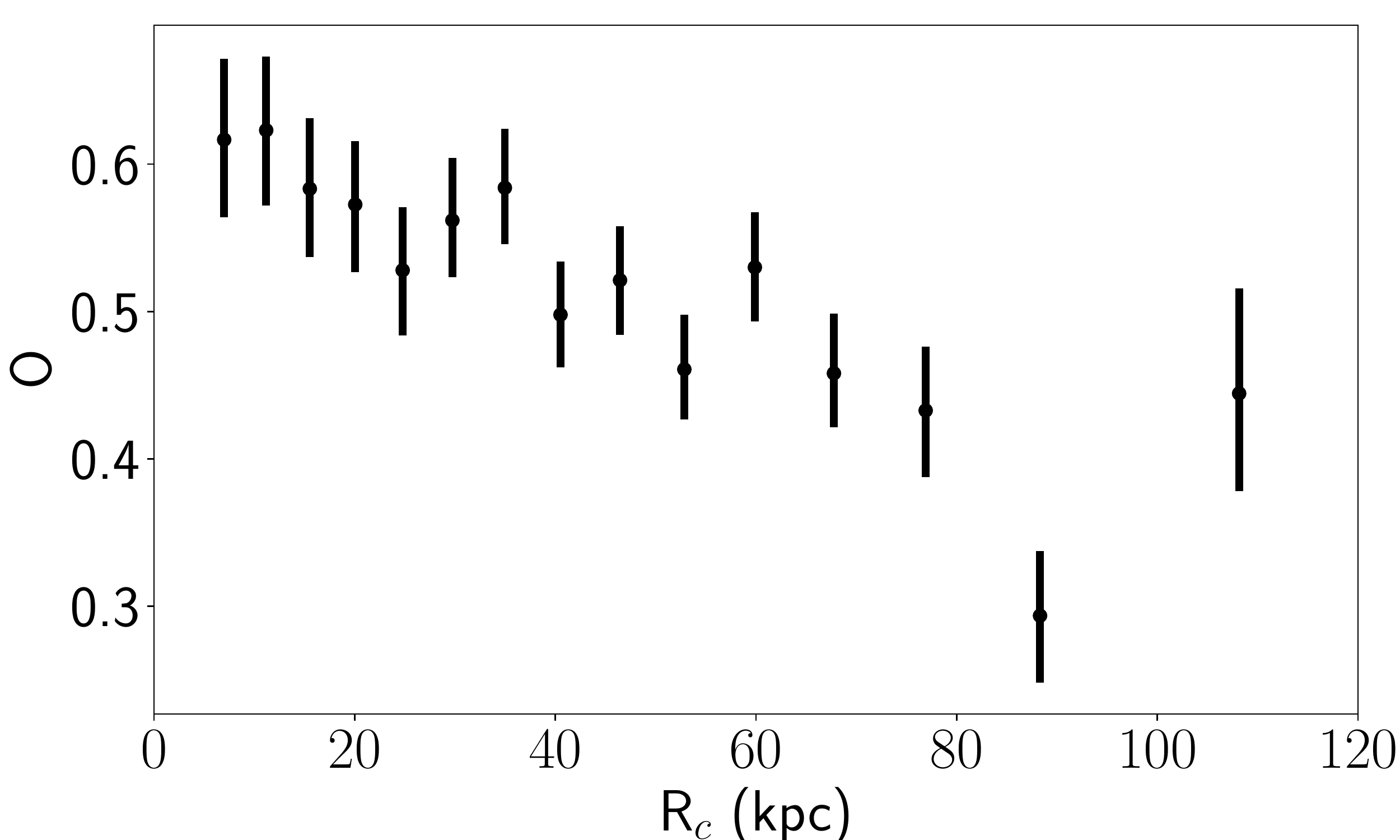}
\includegraphics[width=0.33\textwidth]{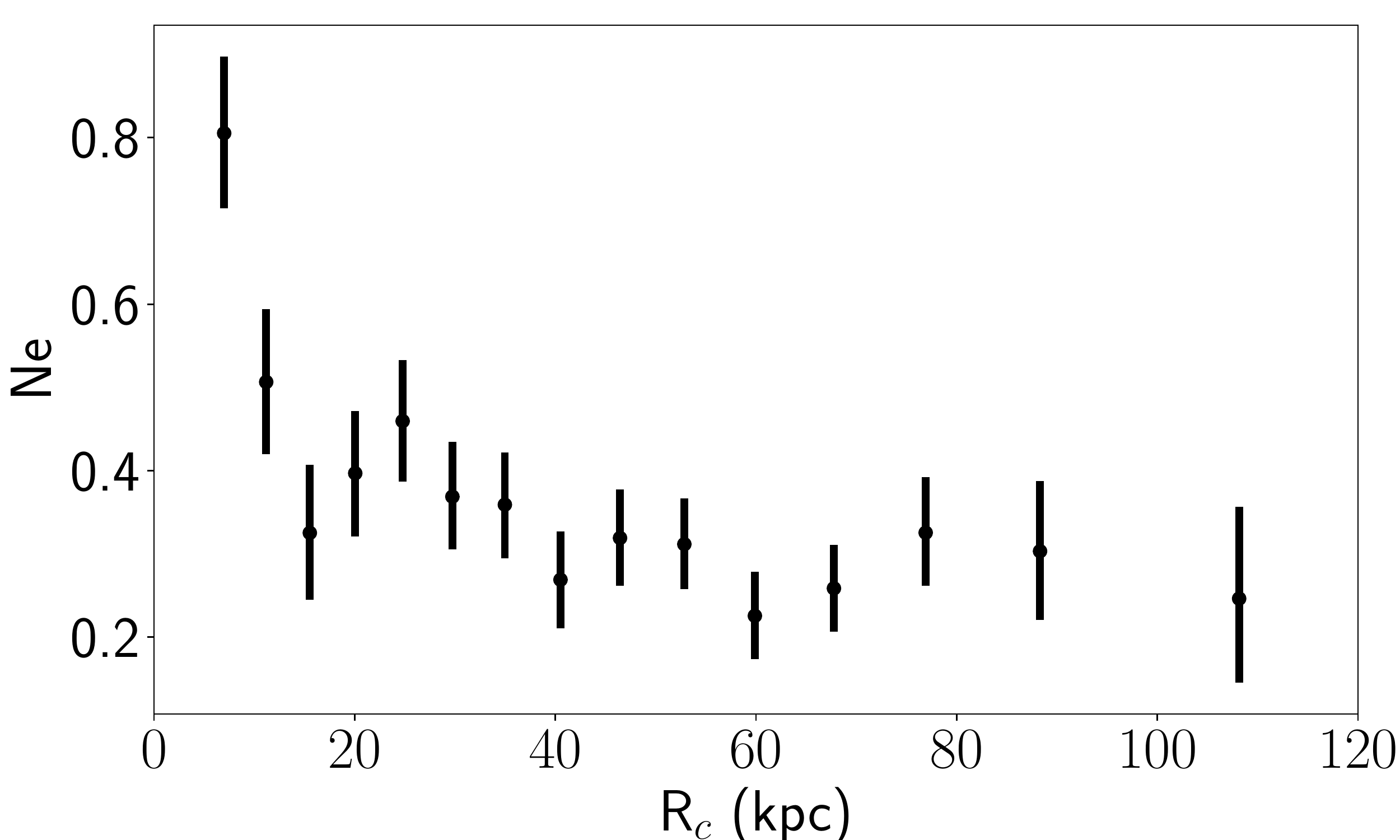} 
\includegraphics[width=0.33\textwidth]{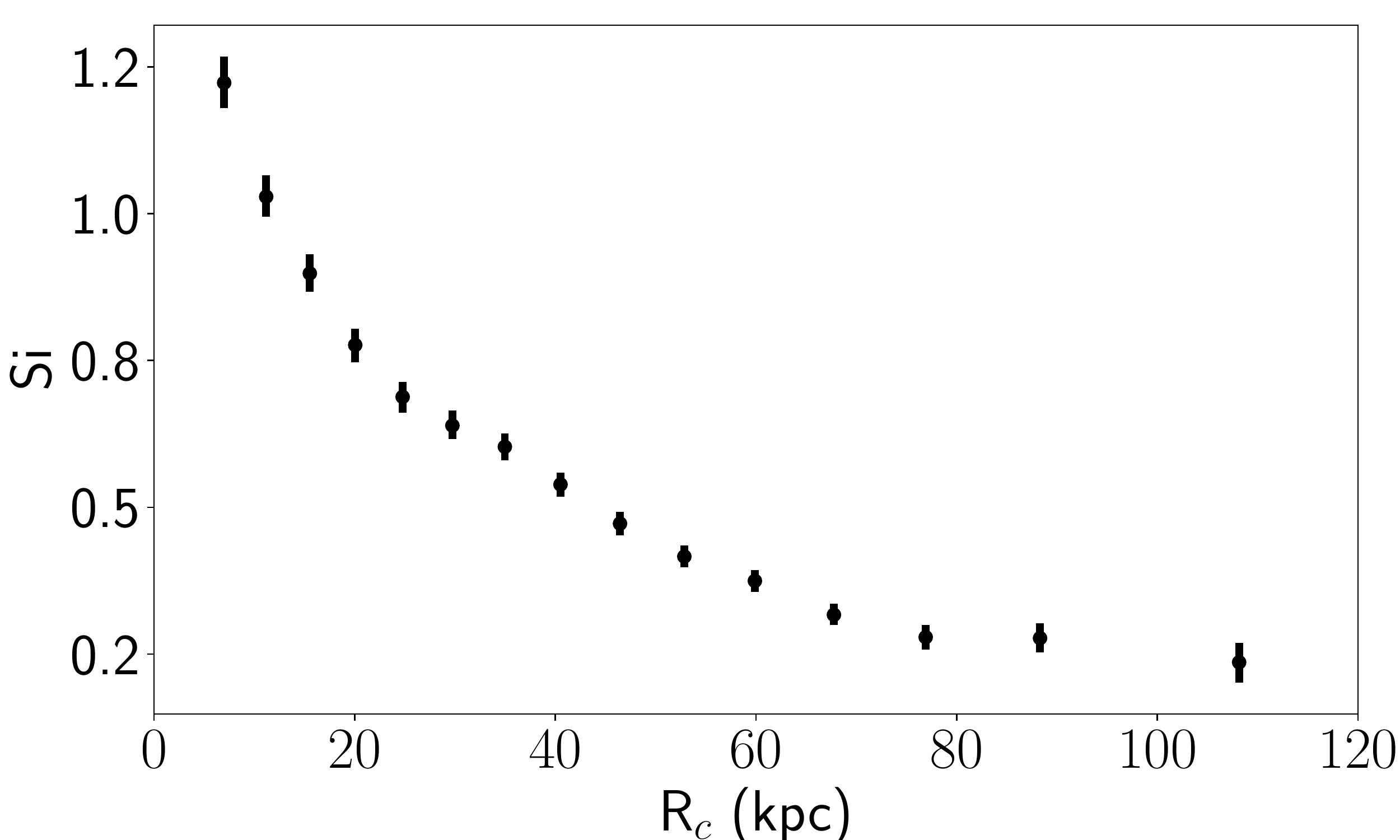}\\
\includegraphics[width=0.33\textwidth]{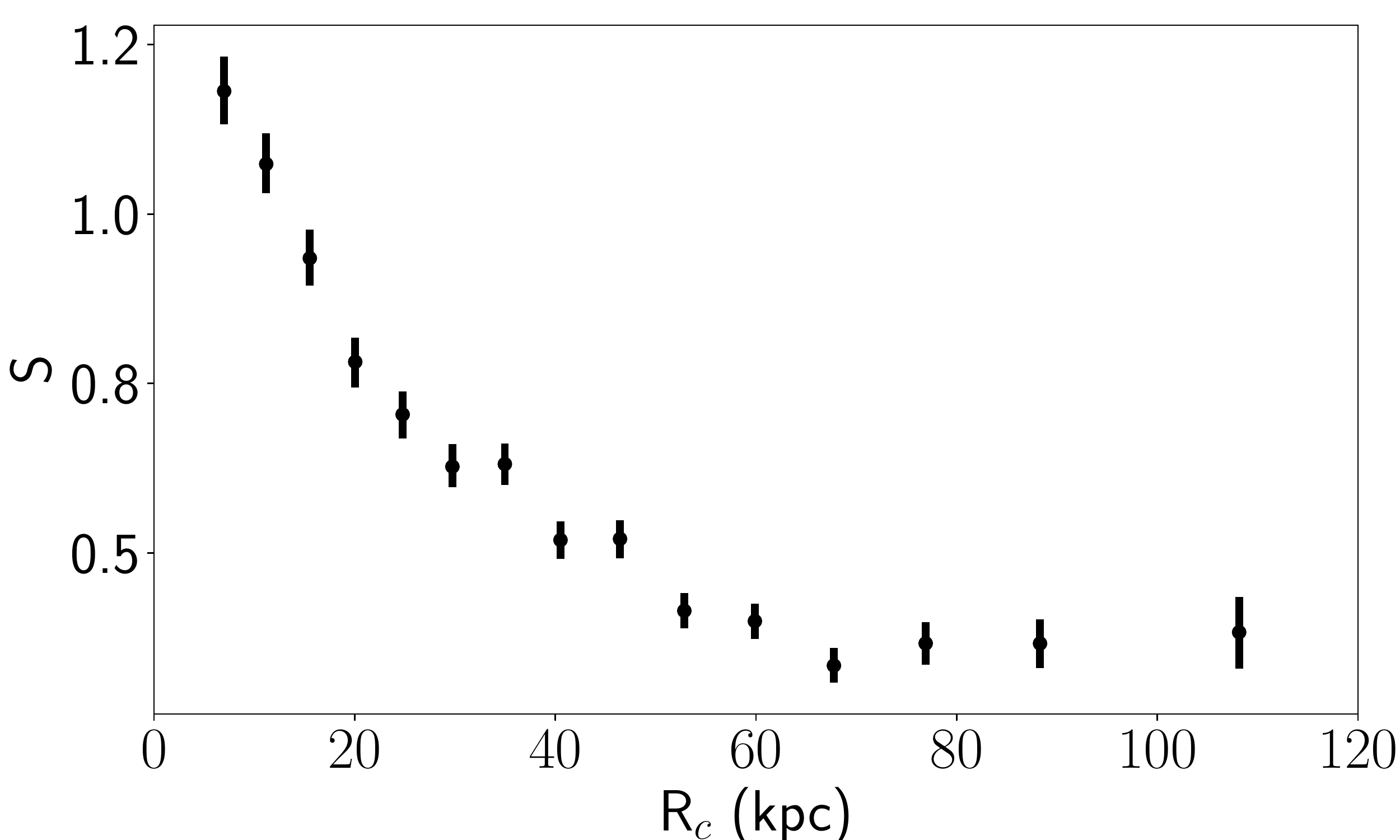}
\includegraphics[width=0.33\textwidth]{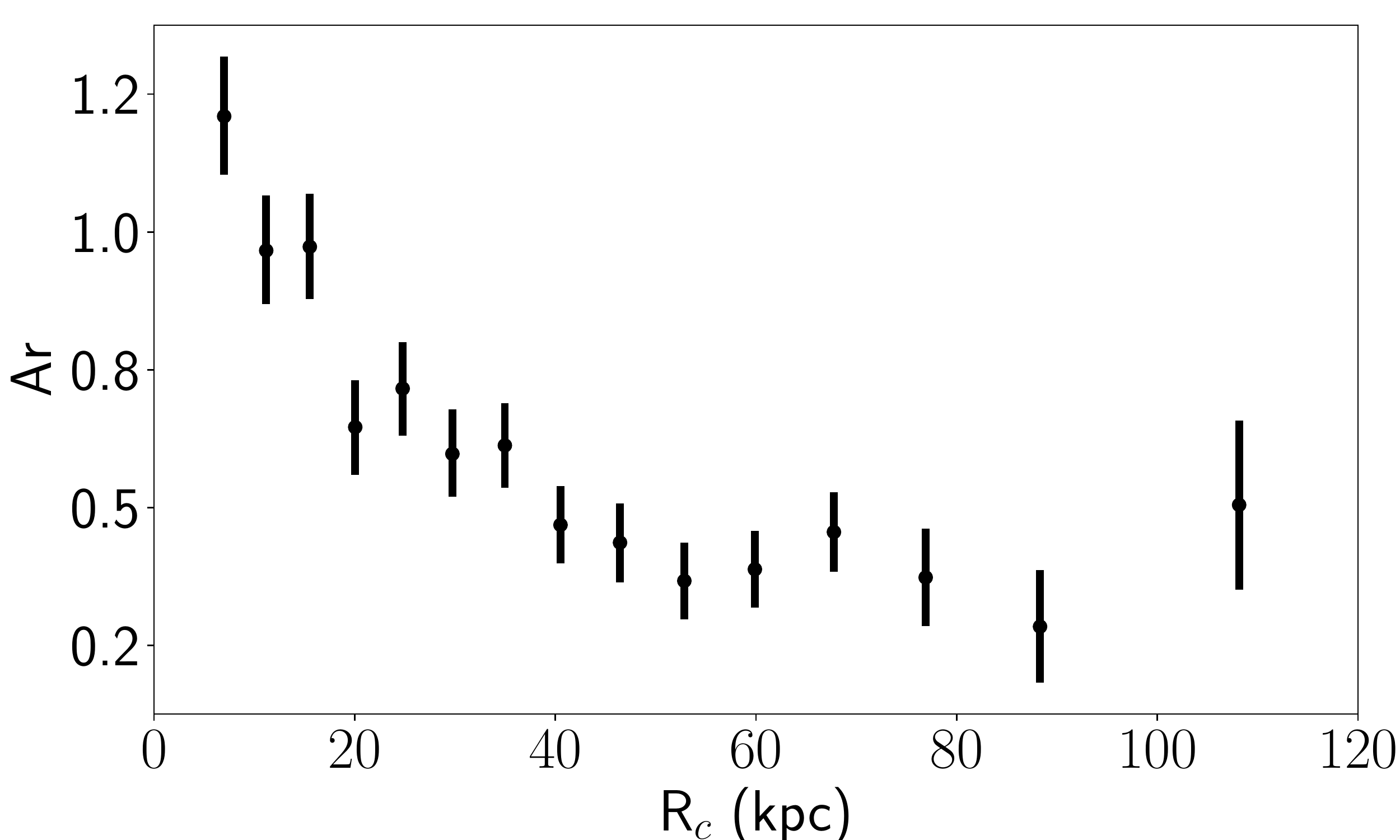}
\includegraphics[width=0.33\textwidth]{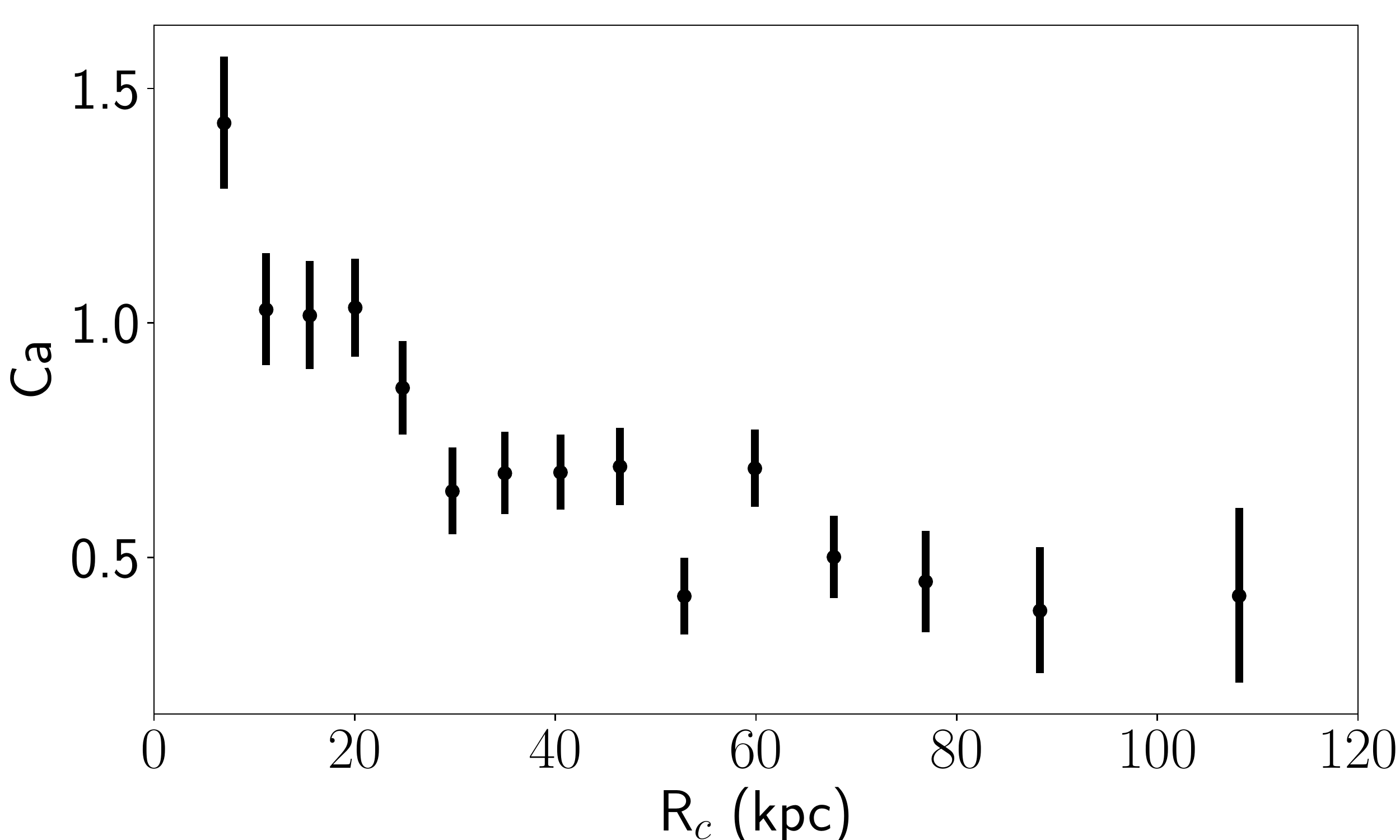}\\
\includegraphics[width=0.33\textwidth]{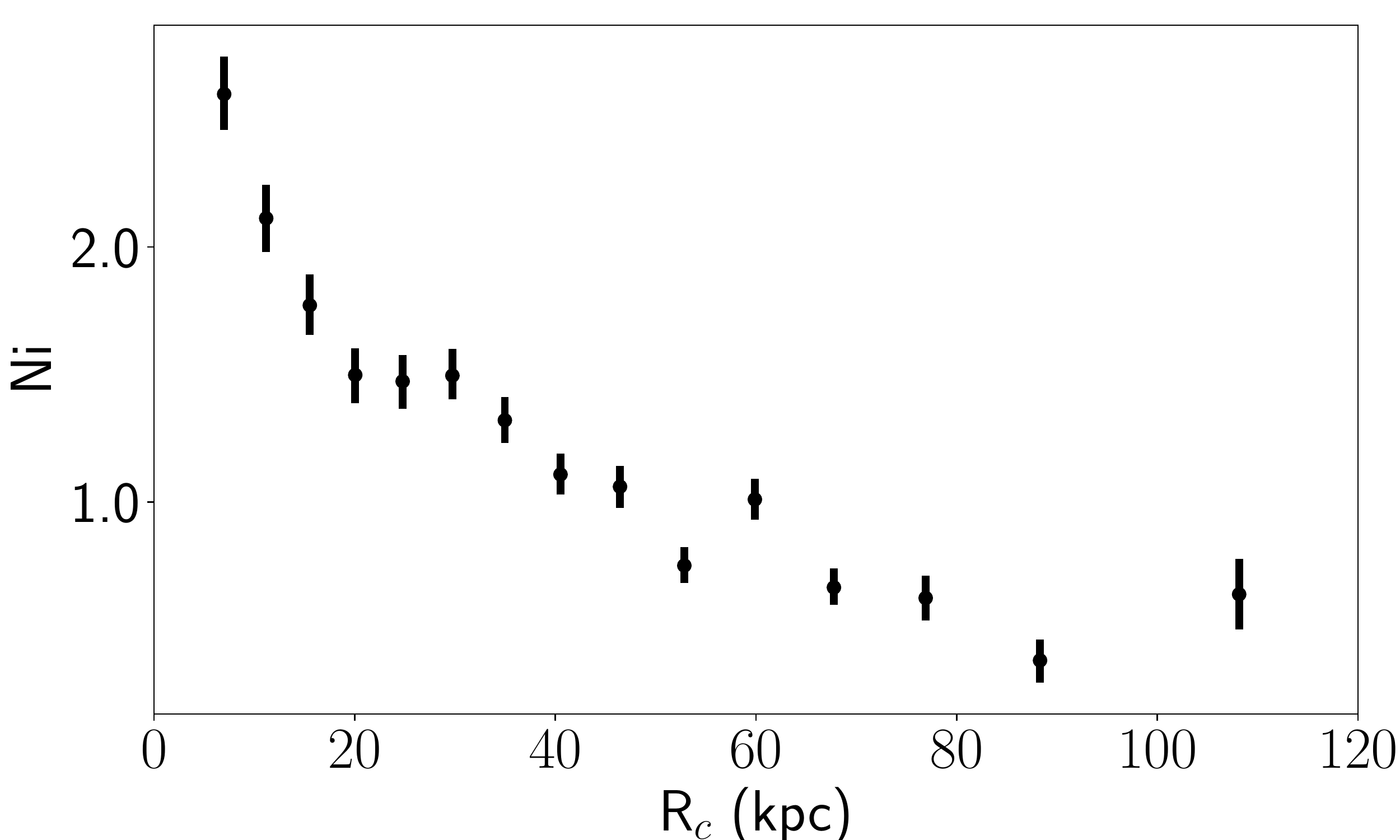} 
\includegraphics[width=0.33\textwidth]{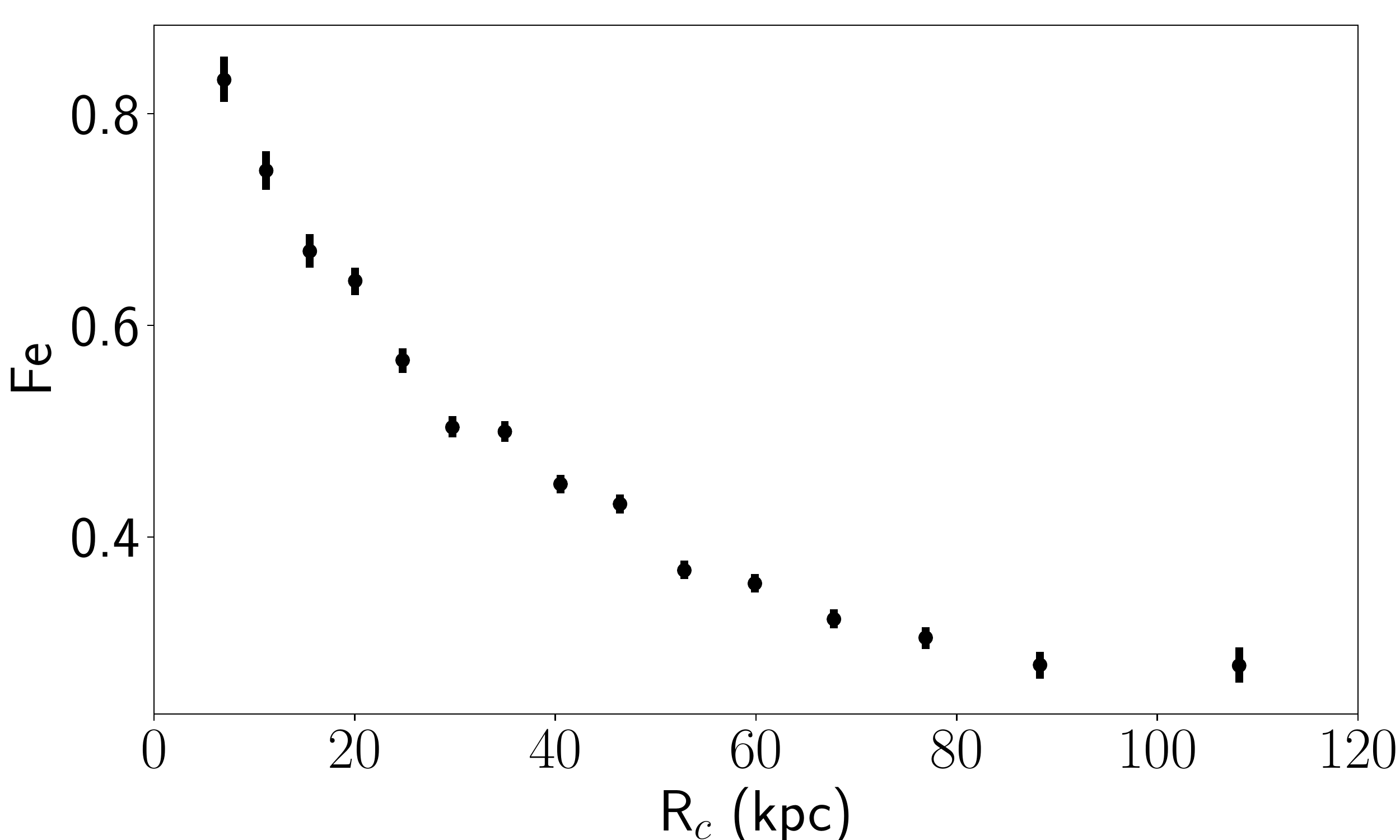}  
\caption{Abundance profiles obtained from the best-fit results.} \label{fig_abund_all} 
\end{figure*}

\begin{figure}    
\centering
\includegraphics[width=0.48\textwidth]{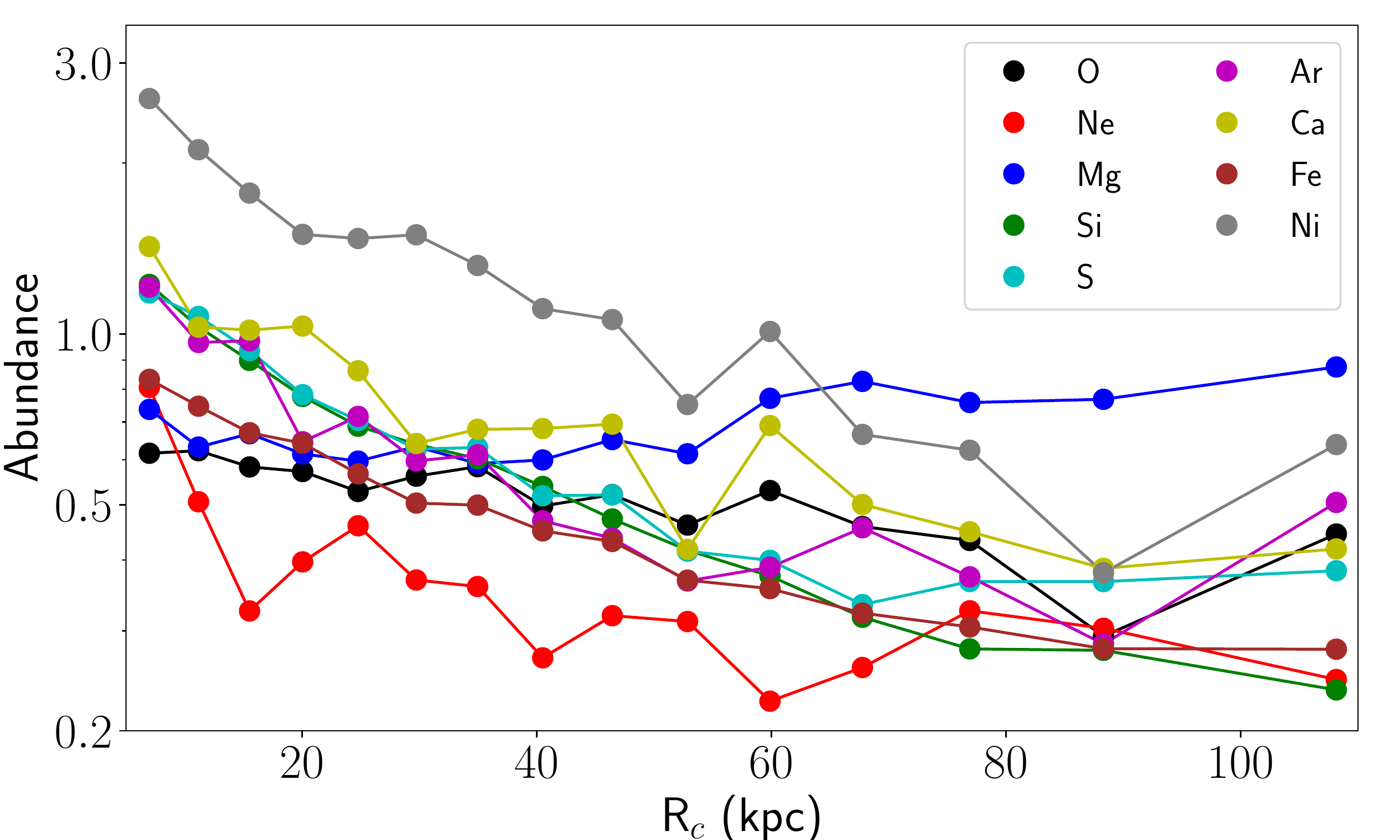}  
\caption{Radial abundances distribution for each element. Uncertainties are not included for illustrative purposes.} \label{fig_abund_together} 
\end{figure}

Figure~\ref{fig_abund_all} shows the best-fit elemental abundances obtained respect to the solar values while Figure~\ref{fig_abund_together} shows the radial abundances distribution in a combined plot. The derived abundances of Si, Ar, S and Fe show steep negative gradients for distances $<60$~kpc before became somewhat flatter. There is a hint for a discontinuity at $\sim30$~kpc. \citet{sim07} discussed the presence of an inner cold front at such distance. Lighter elements such as O and Ne display abrupt drops in their profiles. Interesting, the abundances tend to increase for large radius (at $\sim100$~kpc for O, Ne, Ar). A similar trend at large radii have been found for the other sources \citep{mer17,lak19}. We noted that such increase may be prone to systematics. For example, the Ne line is blended with the Fe-L complex (see Figure~\ref{fig_spectra_example}), therefore the Ne abundance is prone to systematics due the multi-temperature modeling. We noted that O displays a flat profile for distances $<40$~kpc. When comparing with the temperature profile we found that the cooler gas is more metal-rich. Finally, the abundance profiles obtained are in good agreement with previous works covering distances up to $\sim40$~kpc by \citep{wer06,sim10,mil11}. Regarding the cluster core, \citep{dep17} found a ratio O/Fe$=1.0\pm 0.3$ in the central part of the Virgo cluster in the analysis of the CHEmical Enrichment RGS Sample (CHEERS). Similar solar abundance ratios were found in the Perseus cluster \citep{hit17}. However, it is important to note that in this analysis we have excluded the cluster core.

\subsection{ICM chemical enrichment from SN}\label{sec_snr} 
\begin{figure*}    
\centering
\includegraphics[width=0.33\textwidth]{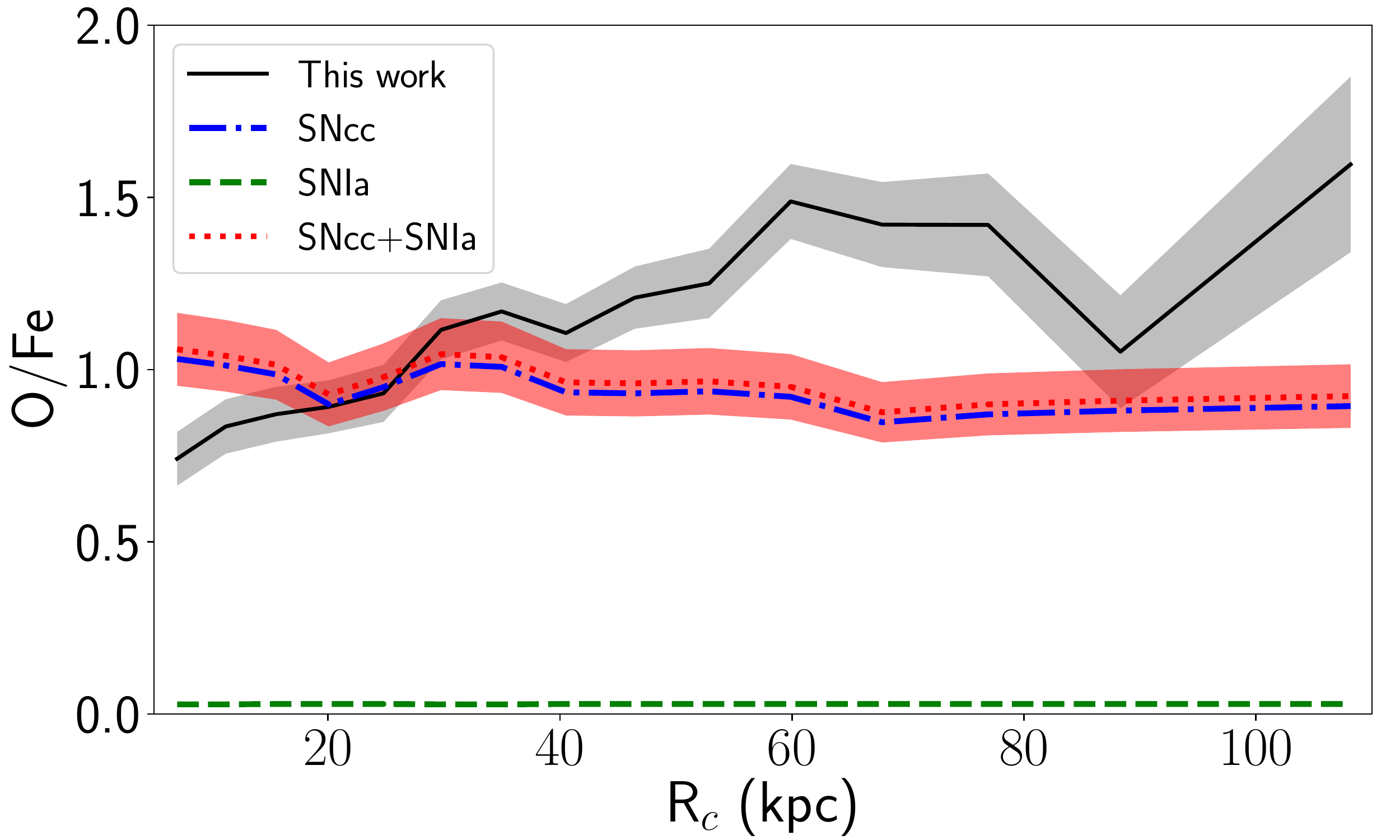}
\includegraphics[width=0.33\textwidth]{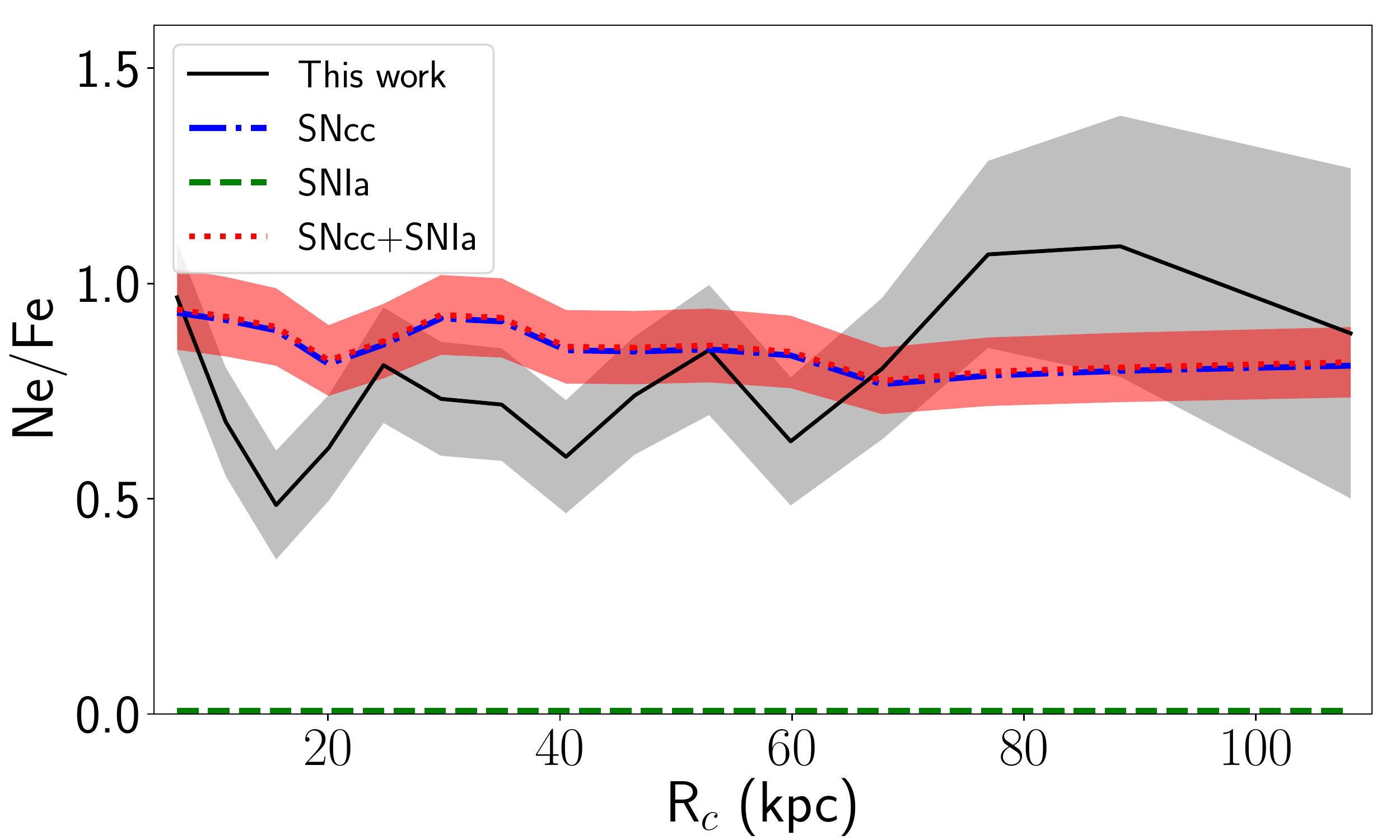} 
\includegraphics[width=0.33\textwidth]{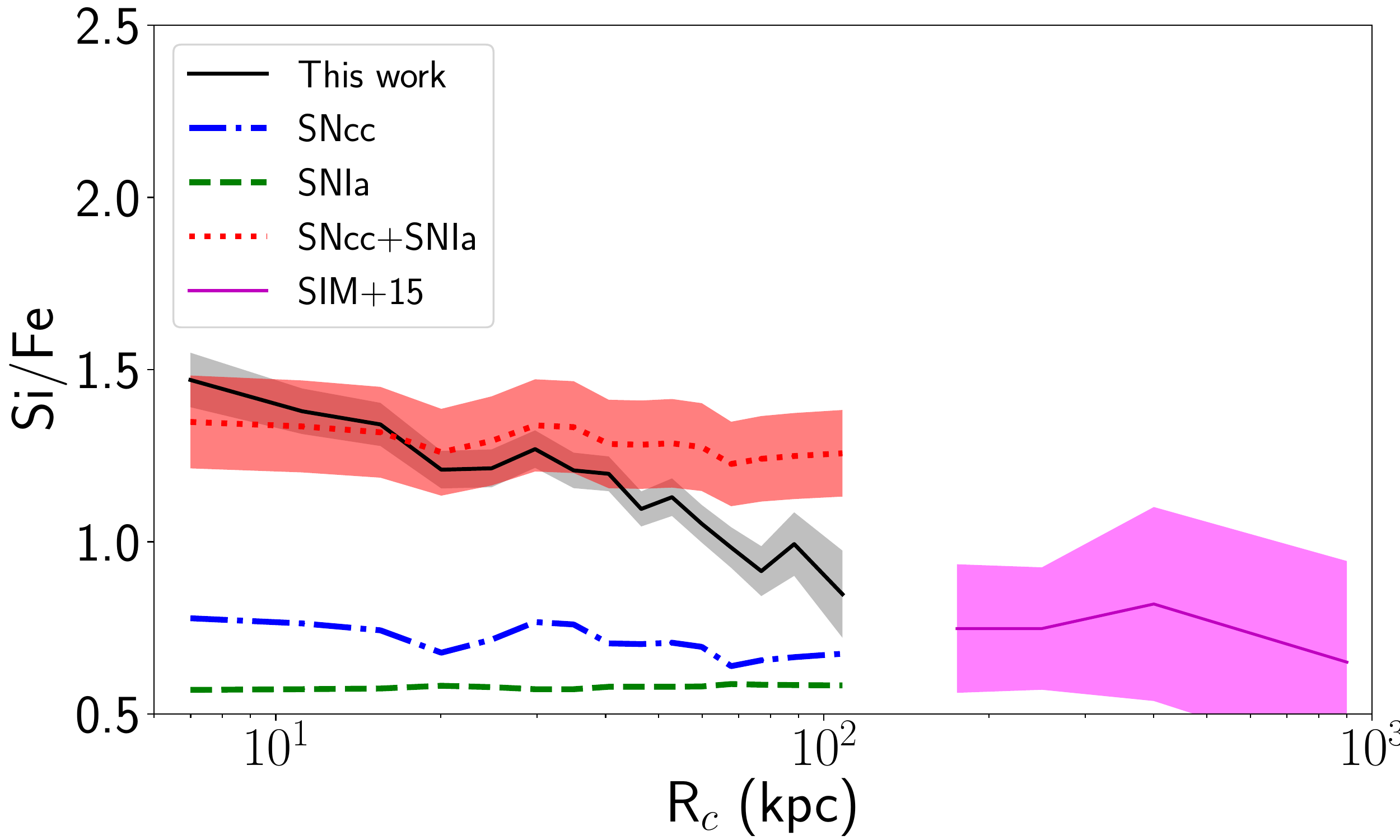} \\
\includegraphics[width=0.33\textwidth]{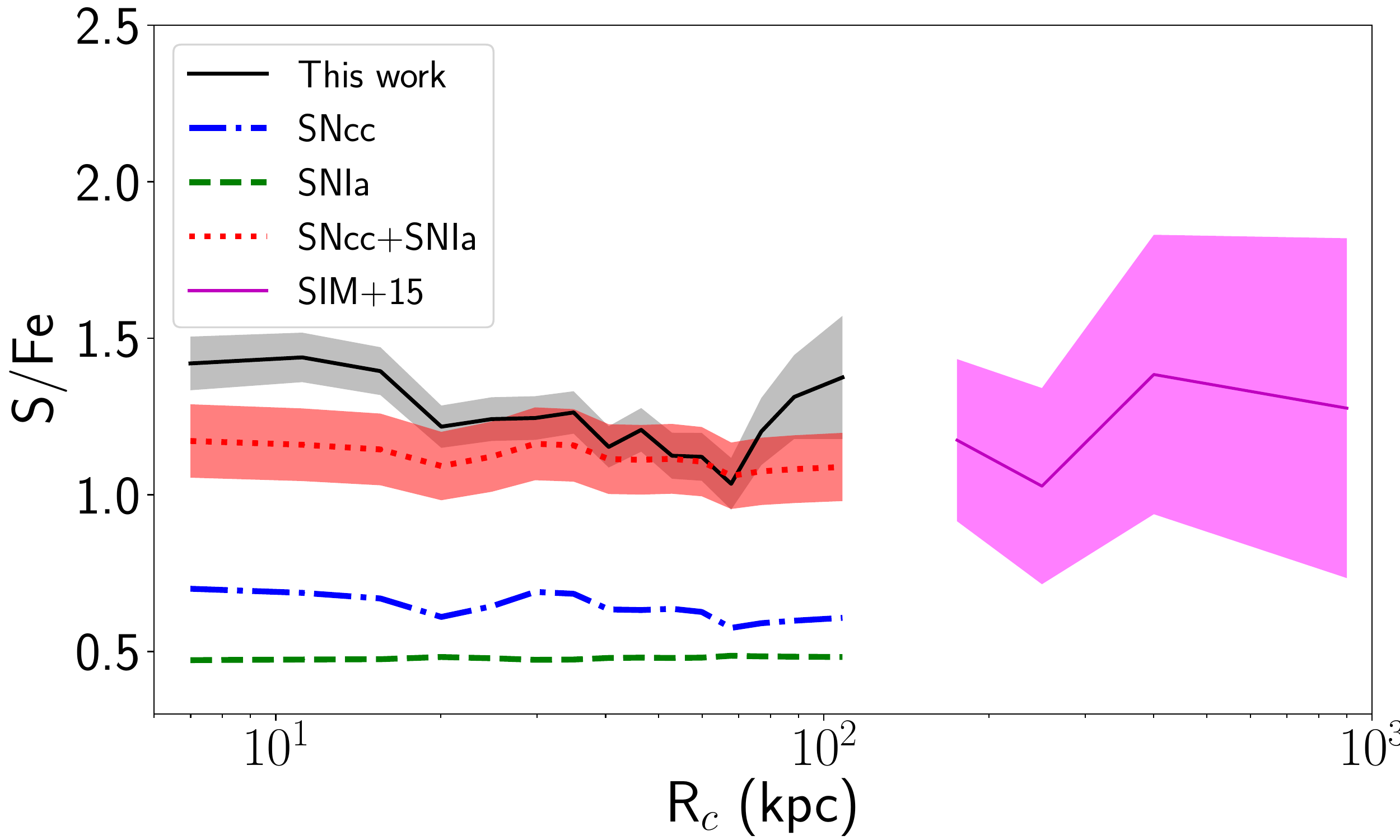} 
\includegraphics[width=0.33\textwidth]{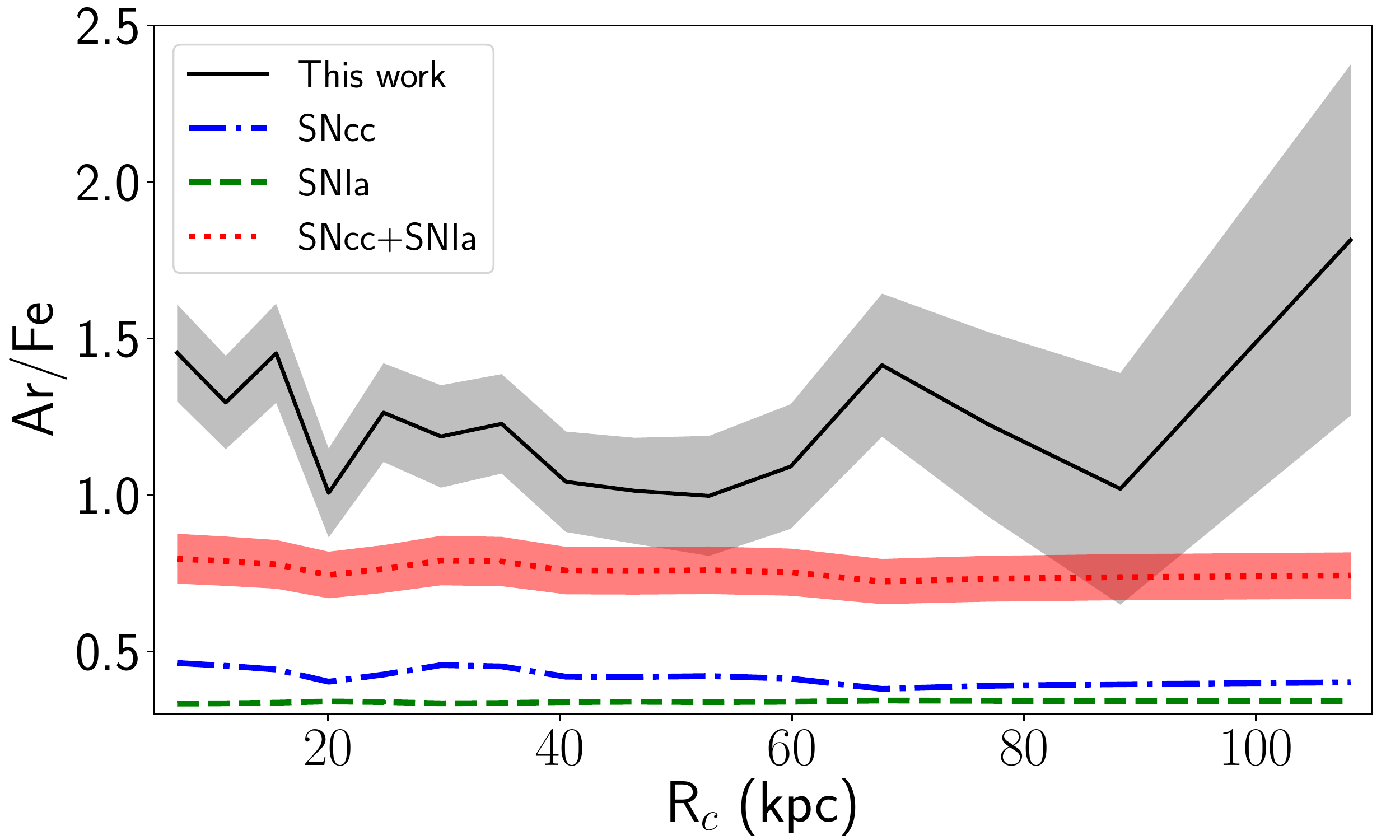}
\includegraphics[width=0.33\textwidth]{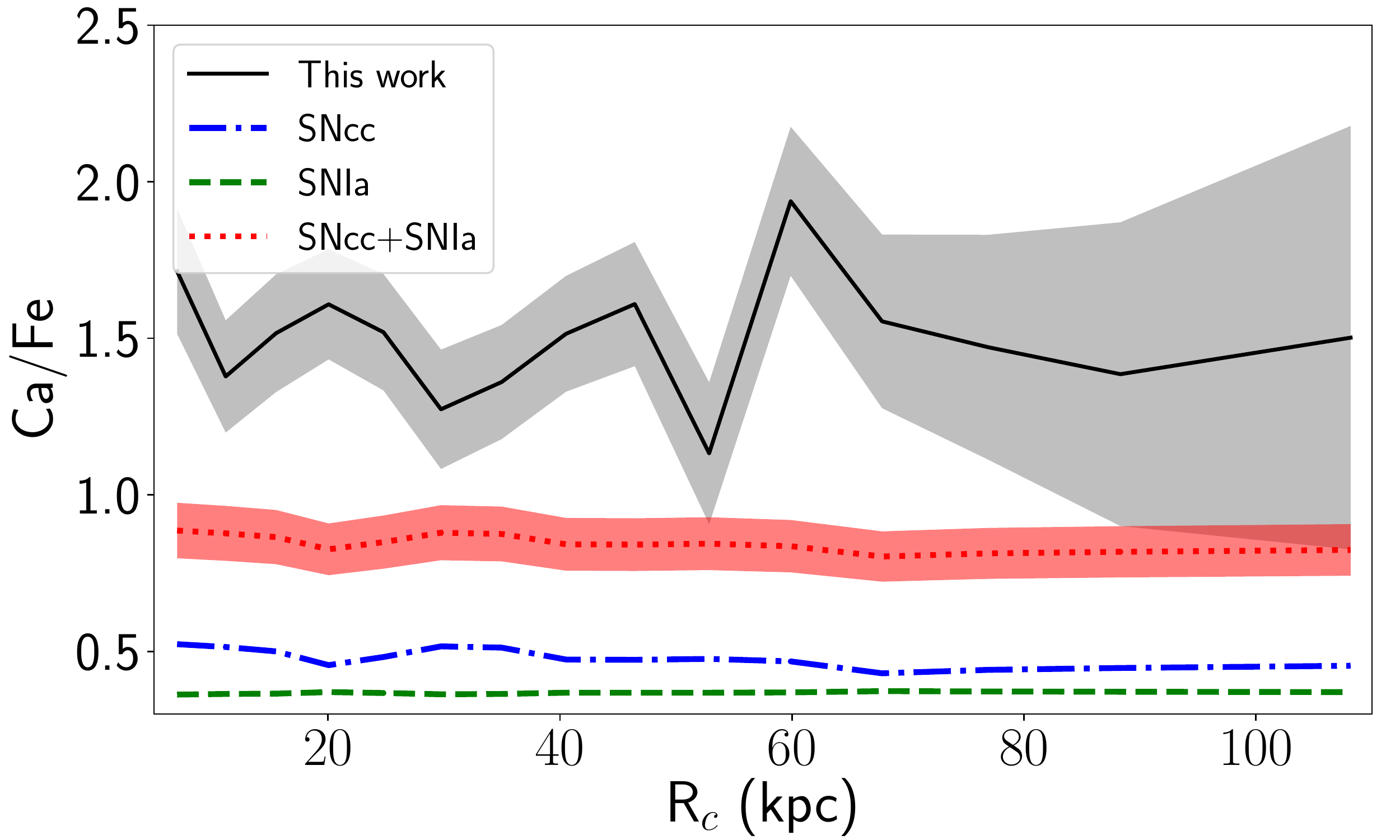}\\
\includegraphics[width=0.33\textwidth]{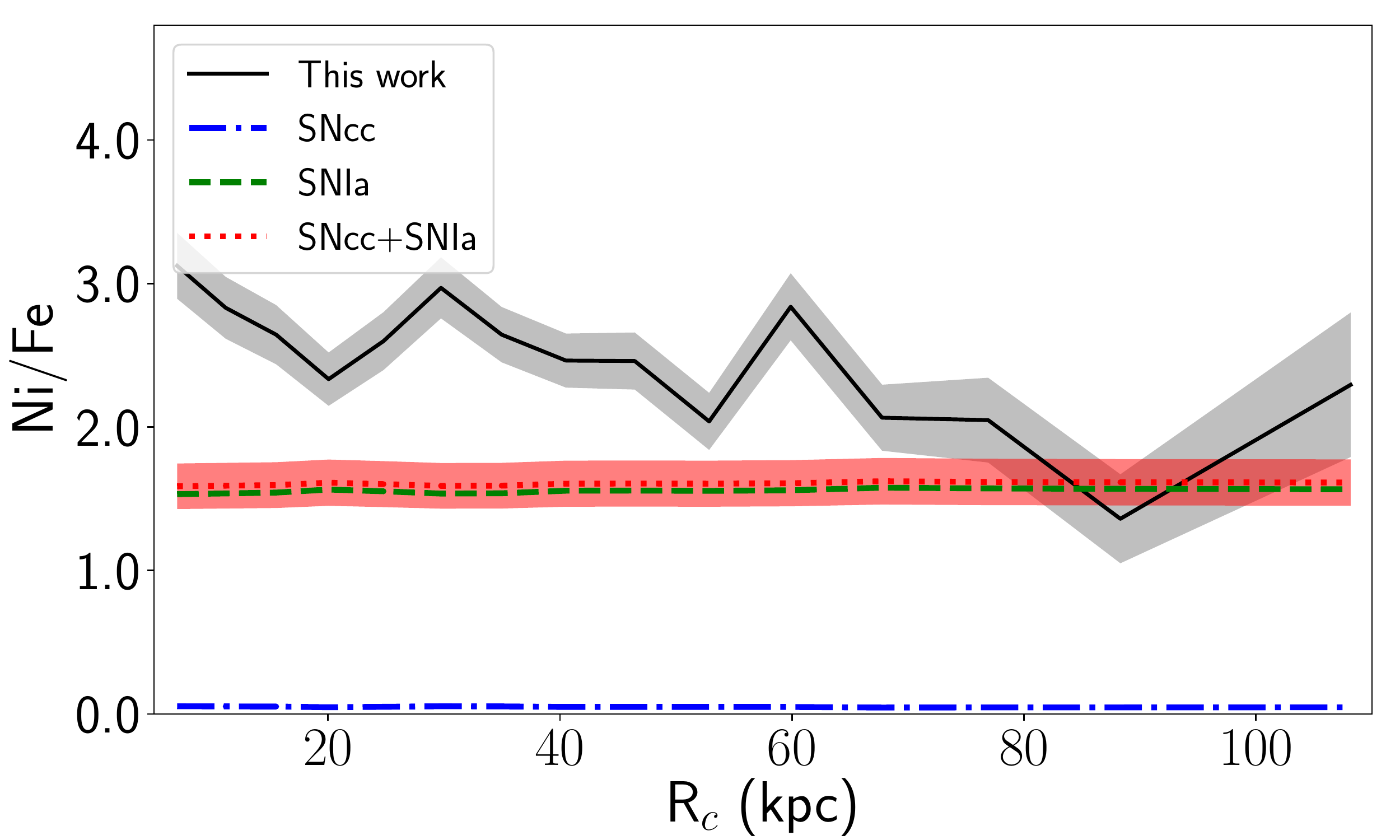} 
\caption{Abundance ratio profiles, relative to Fe. The gray shaded areas indicate the mean values and the 1$\sigma$ errors. The magenta shaded regions are from \citet{sim15}. The SNcc (blue line), SNIa (green line) contribution to the total SN ratio (red shaded area) from the best fit model are included (see Section~\ref{sec_snr}).  } \label{fig_ratios} 
\end{figure*}

Figure~\ref{fig_ratios} shows the X/Fe ratio profiles for all elements measured (gray shaded regions). The uncertainties are lower than $10\%$ for most of the measured values. We also include the Si/Fe and S/Fe ratios obtained by \citet{sim15} from their analysis of {\it Suzaku} observations of the M87 outskirt. We noted that the ratio distribution obtained for those elements in our analysis tend to follow the same shape for large distances. Finally, the ratios obtained are in good agreement with those obtained by \citep{wer06,sim10,mil11}. 

We consider the contribution from different SN yield models to the abundances ratio by using the {\tt SNeRatio} python code developed by \citep{erd21}. This code fits a given set of ICM abundances with a combination of multiple progenitor yield models to calculate the relative contribution that explains the observed data. For SNIa we applied a set of 3D SNIa models in near Chandrasekhar-mass including two different explosion mechanisms: pure deflagration from \citet{fin14} and delayed detonation from \citet{sei13}. Such model have been used in recent enrichment studies \citep{mer17,sim19,mer20}. For the SNcc yields we included models with different initial metallicities from \citet{nom13}. In particular, we included models with metallicity values of Z $=$ (0.0, 0.001, 0.004, 0.008, 0.02, 0.05). Throughout this study, SNcc yields were integrated with Salpeter IMF over the mass range of 10-70 M$_\odot$.  

In order to find the SNe ratio profile for all distances measured we represent the profile with the same model along the radii. We determine the best linear combination of SNIa and SNcc models which better explains all the regions simultaneously by minimizing the sum of their $\chi^{2}$ values in quadrature. We have found that the best reproduction of metals for all regions is achieved by using an initial metallicity of Z $=$ 0.0 for SNcc and a pure deflagration 3D N1600 model for SNIa \citep[see Table~B1 in][]{fin14}. Table~\ref{tab_snr_contribution} shows the SNIa contribution to the total enrichment by this set of models. We have found that the SNIa contribution to the total SN model is $<40\%$ for all radius, following a flat radial distribution. A linear fit gives almost zero slope (4.14$\times$10$^{-4}$ with $\chi^{2}=1.18$) and a constant value of $0.35\pm 0.02$. This support an early enrichment of the ICM scenario, with most of the metals present being produced prior to clustering. Figure~\ref{fig_ratios} shows the X/Fe ratios estimated from this model, which roughly reproduces the observed abundance pattern of these elements. It is worth mentioning that, apart from limitations on the CCD-resolution abundances measurement, current yield calculations are prone to large uncertainties as well. Hence, further effort in improving theoretical models of supernova nucleosynthesis is crucial.

The Si/Fe ratio decrease as we move outwards and can be extrapolated to the values obtained by \citet{sim15}. We emphasize that both elements have the most robustly determined abundances (see Figure~\ref{fig_abund_all}) and given the location of the associated lines, the systematic uncertainties as well as possible biases due to multitemperature structure are minimized in the Si modeling (see Appendix~\ref{sec_apx}). \citet{mil11} measured a radially decreasing Si/Fe ratio in their analysis of {\it Chandra} data for the same system, however, such trend has not been observed elsewhere. In order to reproduce such profile, much larger and unrealistic SNIa contribution would be required. A detailed 2D abundance spatial distribution analysis, which will be done in a forthcoming paper, will help to better understand the nature of the Si/Fe ratio profile.

The Ni/Fe ratio is large and tend to be significantly higher than the solar ratio but is reproduced with the SNIa contribution. However, the Ni-L and K lines are blended into the Fe-L lines and He-like Fe-K$\beta$ line at 7.9\,keV, respectively, and the derived Ni abundances may be affected by systematic uncertainties, as described in the {\it Hitomi} analysis of the Perseus cluster \citep{hit17}. Moreover, the atomic data itself (i.e. the lines and physical processes included in the model) constitutes a crucial ingredient in the measurement of chemical abundances \citep[see for example][]{mer18}. We noted that the Ar/Fe and Ca/Fe are not reproduced satisfactorily by this linear combination models of SN populations. In this sense, \citet{fuk22} have shown that the Ca/Fe ratios show discrepancies of about 50$\%$ between CCD detector, in their analysis of the Centaurus cluster core. Both, Ar/Fe and Ca/Fe profiles are quite similar even for large distances, even though Ca can be easily trapped into dust grains. We tested a new linear combination of SN models after excluding the Ar and Ca elements and we have found that the best-fit model is the same as the one obtained before. 

We also analyzed the SNIa contribution to the abundance ratio profiles without including O/Fe ratios, given the impact of the Galactic absorption in the O abundance determination (see Appendix~\ref{sec_multi_fit}). Figure~\ref{fig_ratios2} shows the results. In this case, we also found a flat radial distribution of SNIa ratio over the total cluster enrichment. The Si/Fe and Ni/Fe are better constrained while the Ca/Fe and Ar/Fe ratios are not reproduced satisfactorily.

\begin{table}
%\scriptsize 
\caption{\label{tab_snr_contribution}SNIa contributions to the total chemical enrichment. }
\centering
\begin{tabular}{ccccccc}
\\
Radius  & SNIa & Radius & SNIa \\
 (kpc) & & (kpc)   & \\ 
\hline 
6.99 & $35\pm 5 \% $ &46.45& $38\pm 7 \% $\\ 
11.18 & $36\pm 6 \% $&52.86& $38\pm 8 \% $\\ 
15.53 &$36\pm 6 \% $ &59.89&$38\pm 8 \% $ \\
20.05&$38\pm 7 \% $&67.77&$39\pm 9 \% $ \\ 
24.78 &$38\pm 7 \% $ &76.91& $40\pm 7 \% $\\ 
29.74 &$36\pm 7 \% $ &88.31& $38\pm 6 \% $\\
34.97& $36\pm 7 \% $&108.15&$40\pm 5 \% $ \\
40.51& $37\pm 8 \% $\\
\\ 
 \hline
\end{tabular}
\end{table}

\begin{figure*}    
\centering
\includegraphics[width=0.33\textwidth]{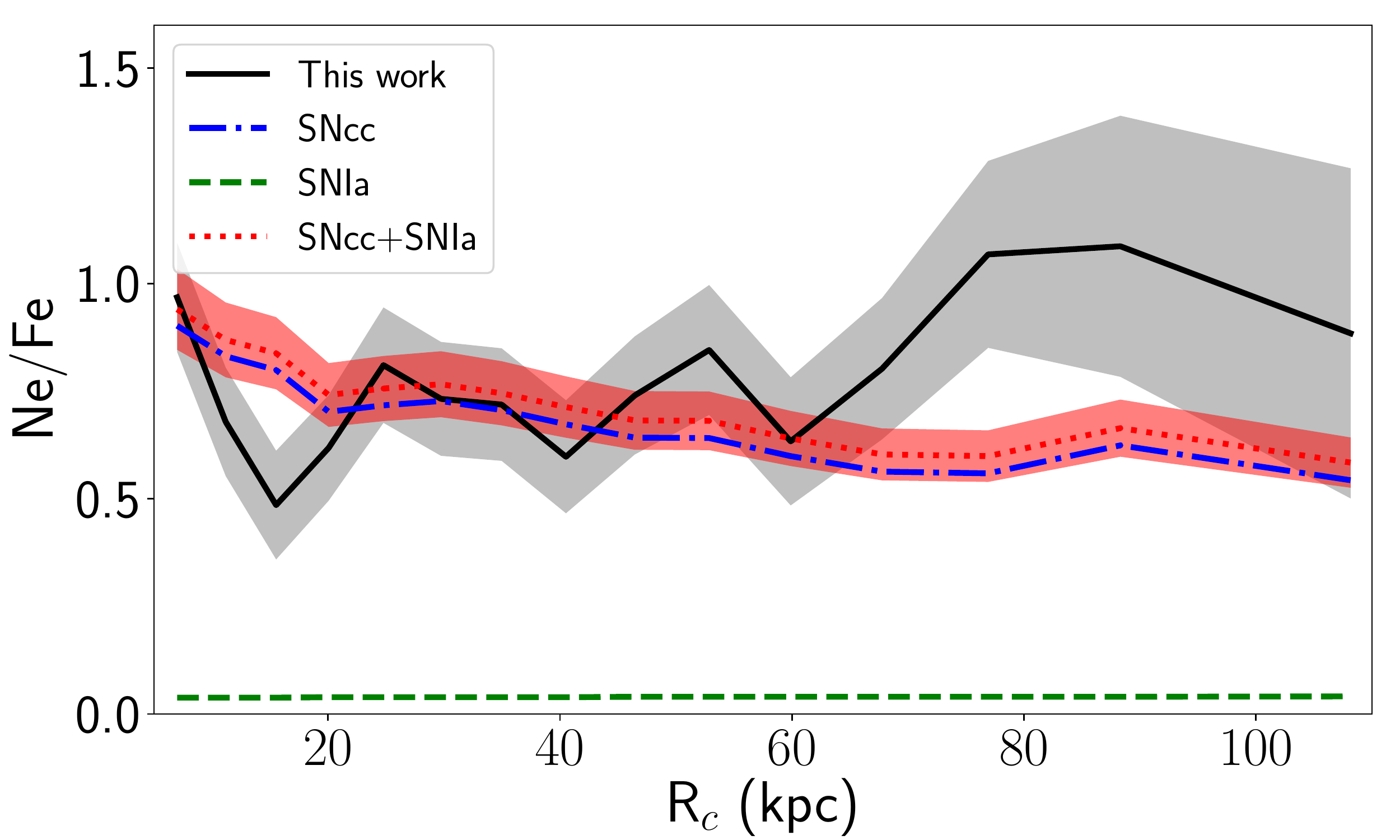} 
\includegraphics[width=0.33\textwidth]{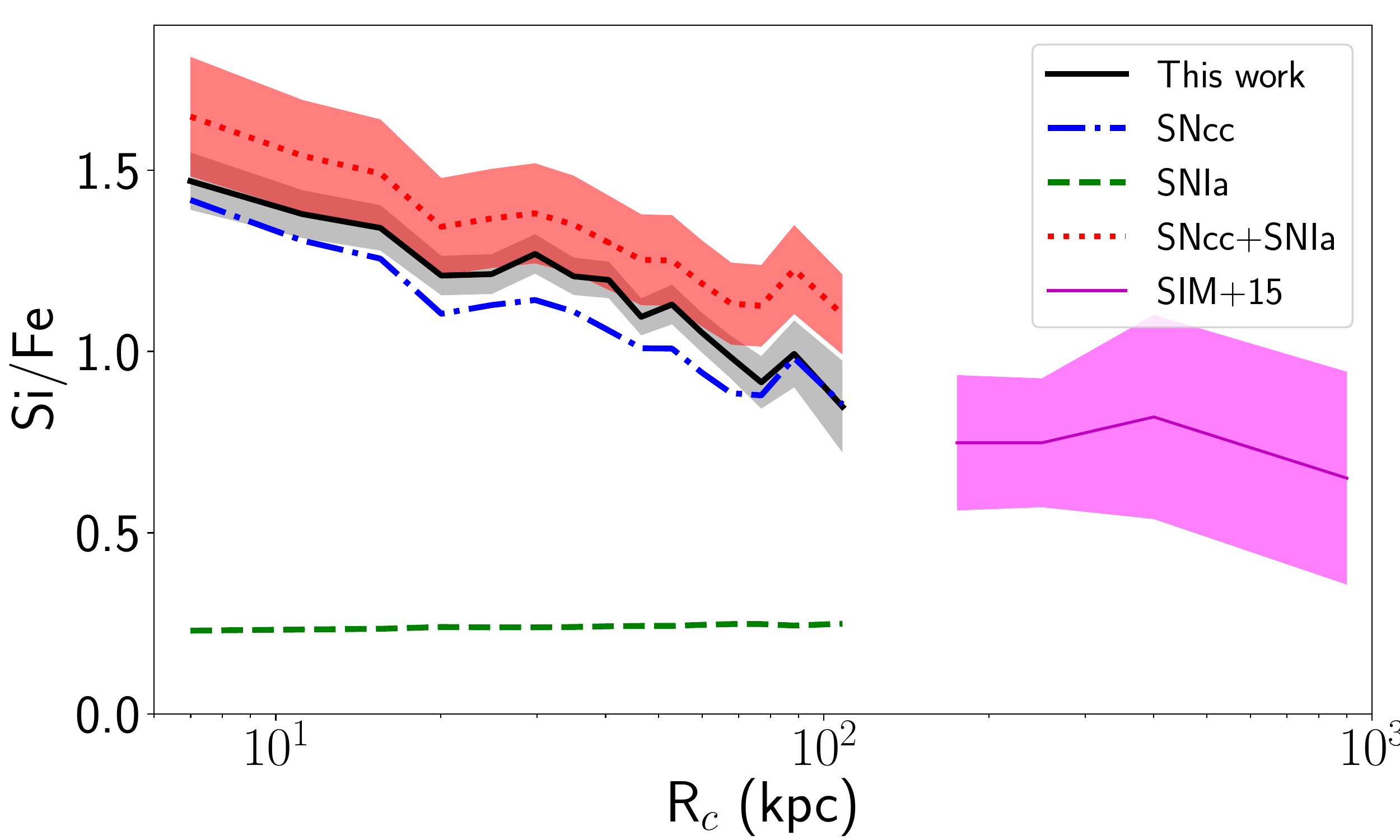} 
\includegraphics[width=0.33\textwidth]{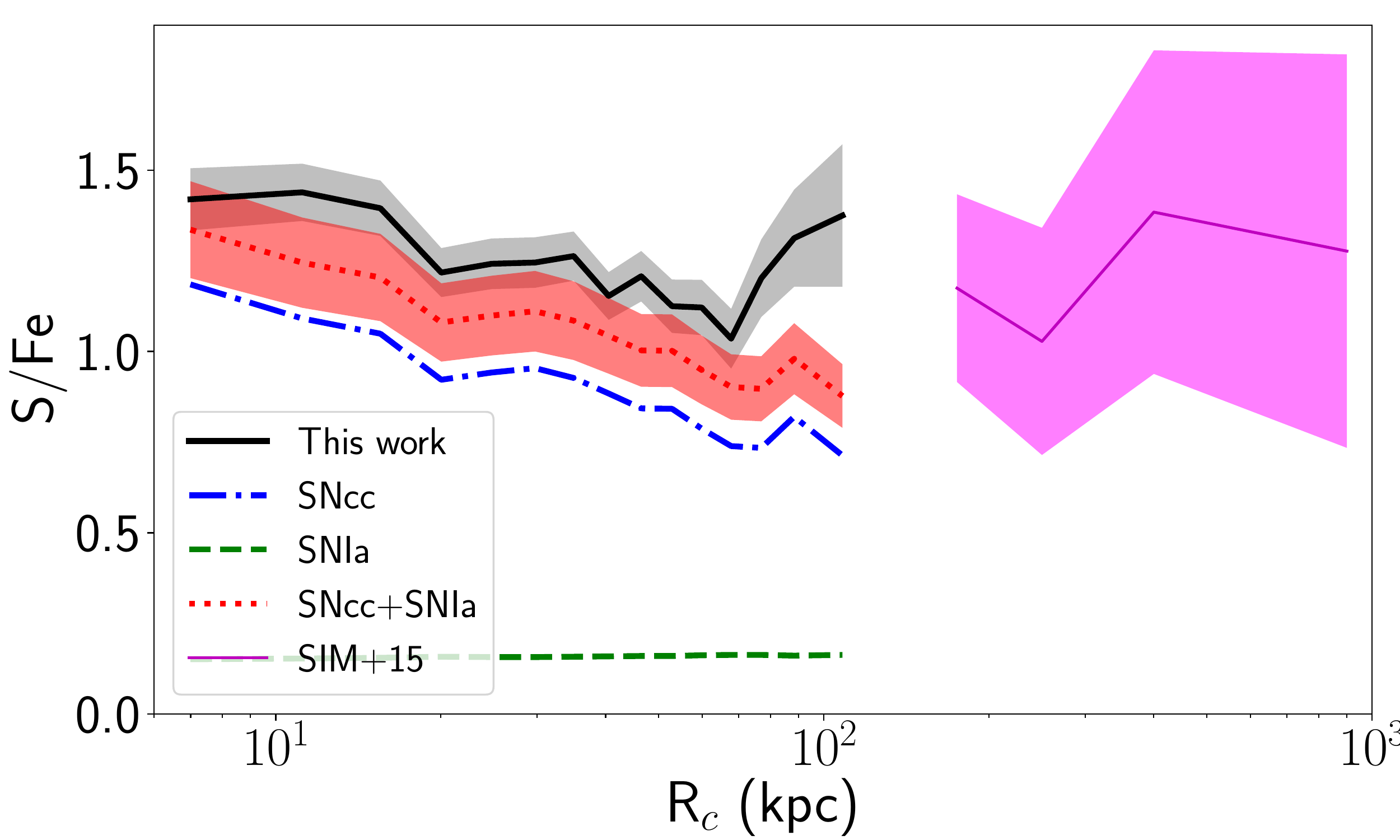}\\
\includegraphics[width=0.33\textwidth]{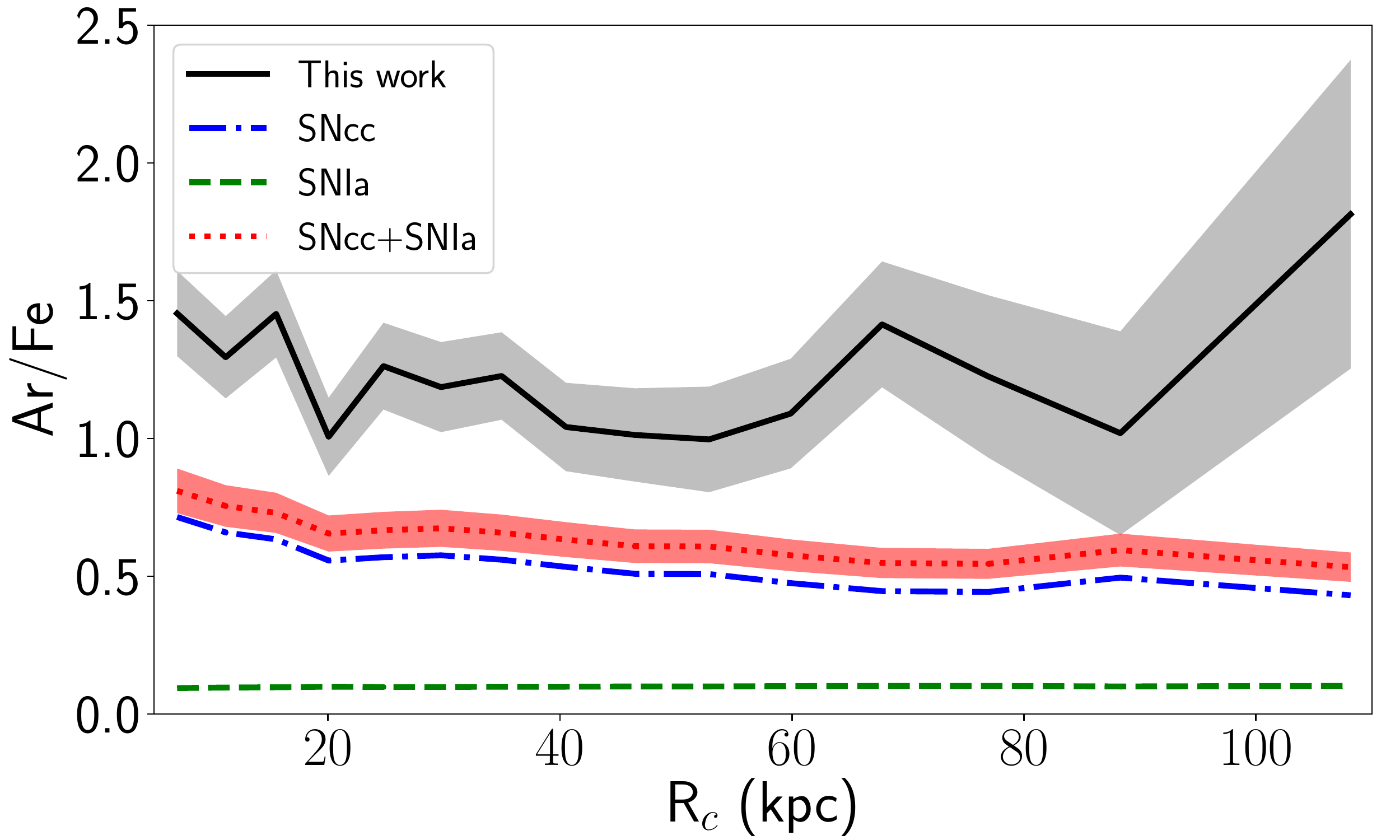}
\includegraphics[width=0.33\textwidth]{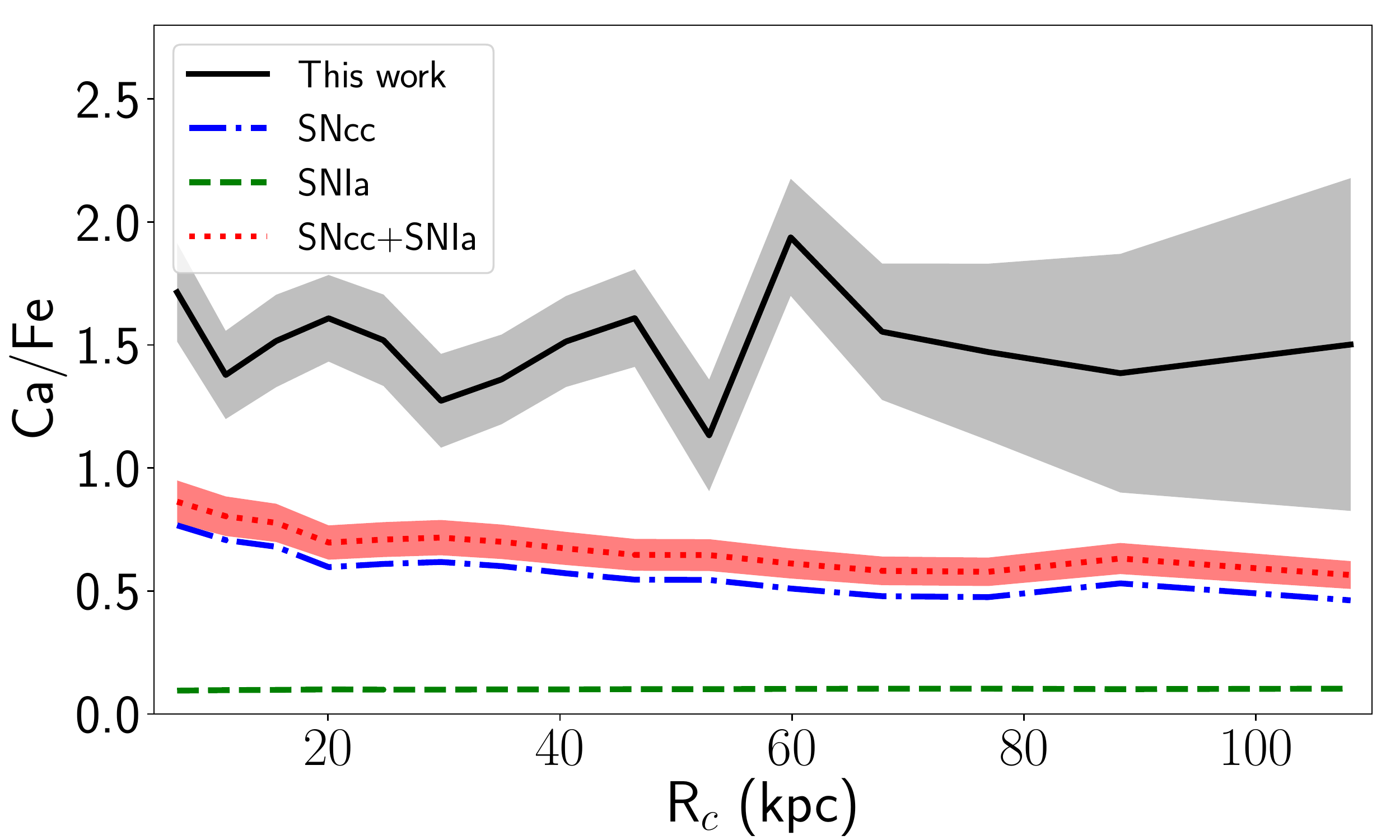}
\includegraphics[width=0.33\textwidth]{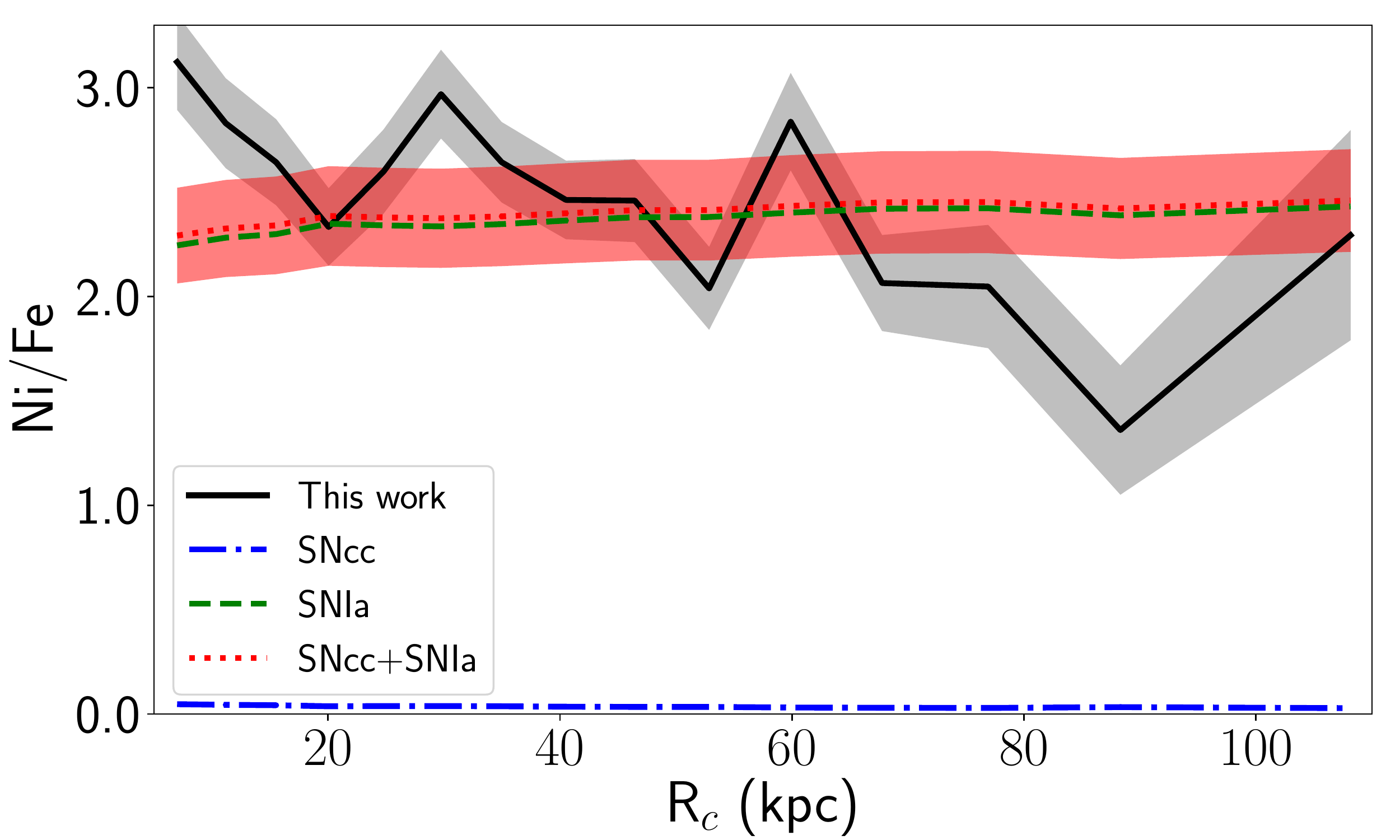} 
\caption{Abundance ratio profiles, relative to Fe, without including O. The gray shaded areas indicate the mean values and the 1$\sigma$ errors. The magenta shaded regions are from \citet{sim15}. The SNcc (blue line), SNIa (green line) contribution to the total SN ratio (red shaded area) from the best fit model are included (see Section~\ref{sec_snr}).  } \label{fig_ratios2} 
\end{figure*}

\section{Conclusions and summary}\label{sec_con} 
We present an analysis of the radial distribution of elements in the ICM within the Virgo cluster using {\it XMM-Newton} EPIC-pn observations. This work is an extension of the results found by \citet{gat22a} but including the soft energy band ($<4$~keV) and focuses on the abundance radial distribution. In this Section we briefly summarize our findings.

\begin{enumerate}
\item We found a velocities trend similar to that found by \citet{gat22a} but with lower values, most likely due to the inclusion of the spectra energy soft band, which allows a better constrain of metallicities and temperatures. 
\item For distances $\sim 90$~kpc the best-fit model is close to a single-temperature component. We have found discontinuities in temperature around $\sim30$~kpc and $\sim90$~kpc, which correspond to the radius of the cold fronts. The material with higher temperatures ($>2.3$~keV) tends to have larger velocities. When comparing the abundance and temperature profiles we found that the cooler gas is more metal-rich. 
\item We modeled the elemental X/Fe ratio profiles for elements O, Ne, Si, Ar, S, Ca, Ni with a linear combination of SNIa and SNcc models. We found that the best fit-model correspond to a pure deflagration 3D model for SNIa and an initial metallicity of Z $=$ 0.0 for SNcc. This model roughly reproduces the observed abundance patterns of Ne/Fe, Si/Fe and Ni/Fe. The Ar/Fe and Ca/Fe profiles  are not reproduced.
\item We found a flat radial distribution of SNIa ratio over the total cluster enrichment. This support an early enrichment of the ICM scenario, with most of the metals present being produced prior to clustering.
\end{enumerate}

The present work will be followed by a detailed study of the 2D spatial distribution of elemental abundances.

\section{Acknowledgements} 
 The authors thank K. Erdim for provide a python version of the {\tt SNeRatio}. This work was supported by the Deutsche Zentrum f\"ur Luft- und Raumfahrt (DLR) under the Verbundforschung programme (Messung von Schwapp-, Verschmelzungs- und R\"uckkopplungsgeschwindigkeiten in Galaxienhaufen). This work is based on observations obtained with XMM-Newton, an ESA science mission with instruments and contributions directly funded by ESA Member States and NASA. H. Russell acknowledges support from an STFC Ernest Rutherford Fellowship and an Anne McLaren Fellowship.
\subsection*{Data availability}
The observations analyzed in this article are available in the {\it XMM-Newton} Science Archive (XSA\footnote{\url{http://xmm.esac.esa.int/xsa/}}).

\bibliographystyle{mnras}
 \newcommand{\noop}[1]{}

\appendix\label{sec_apx}

\section{Spectral models}\label{sec_multi_fit}
\begin{figure}    
\centering
\includegraphics[width=0.48\textwidth]{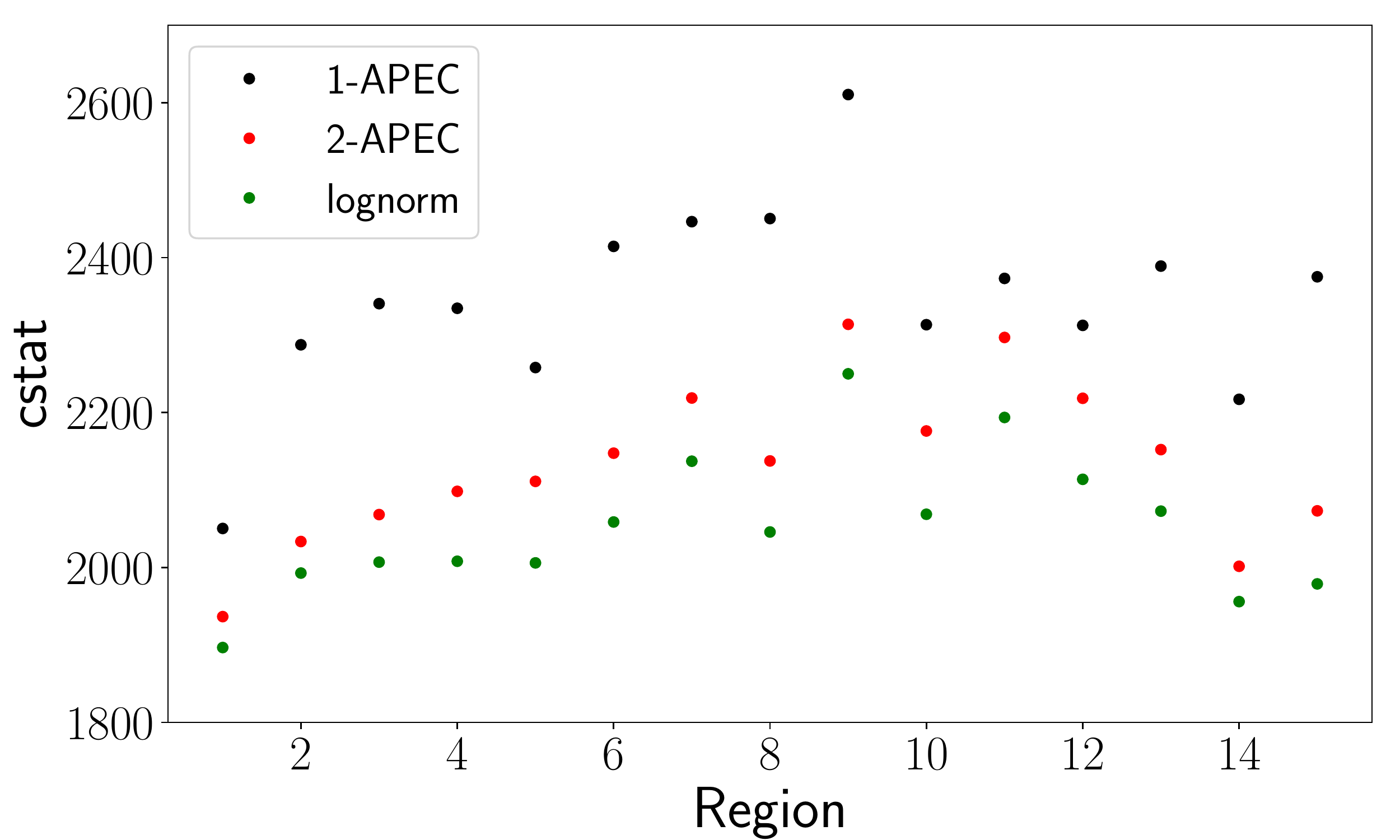} 
\caption{Best-fit cash statistic obtained with the 1-{\tt apec}, 2-{\tt apec} and {\tt lognorm} models.  } \label{cstat_comparison} 
\end{figure}

\begin{figure*}    
\centering
\includegraphics[width=0.48\textwidth]{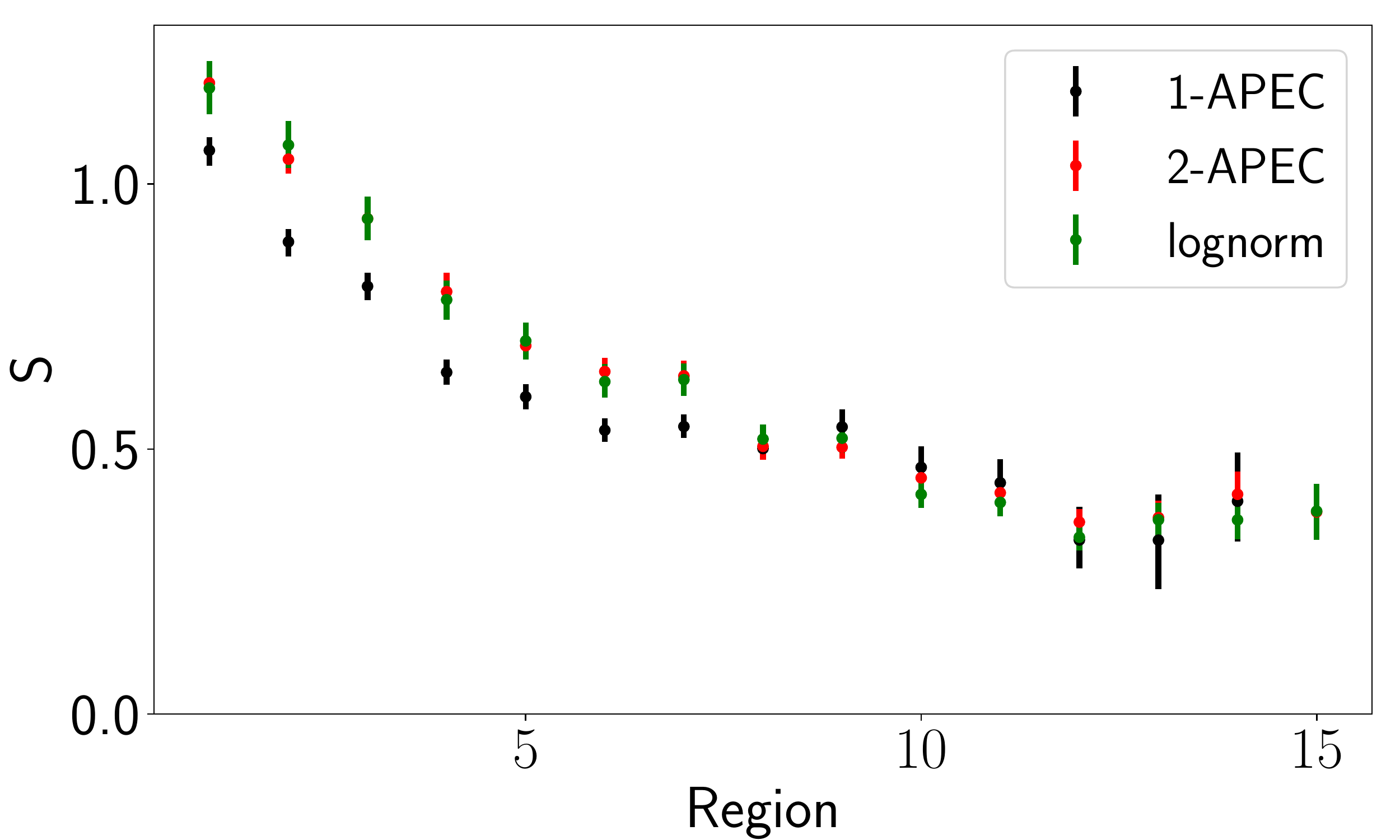} 
\includegraphics[width=0.48\textwidth]{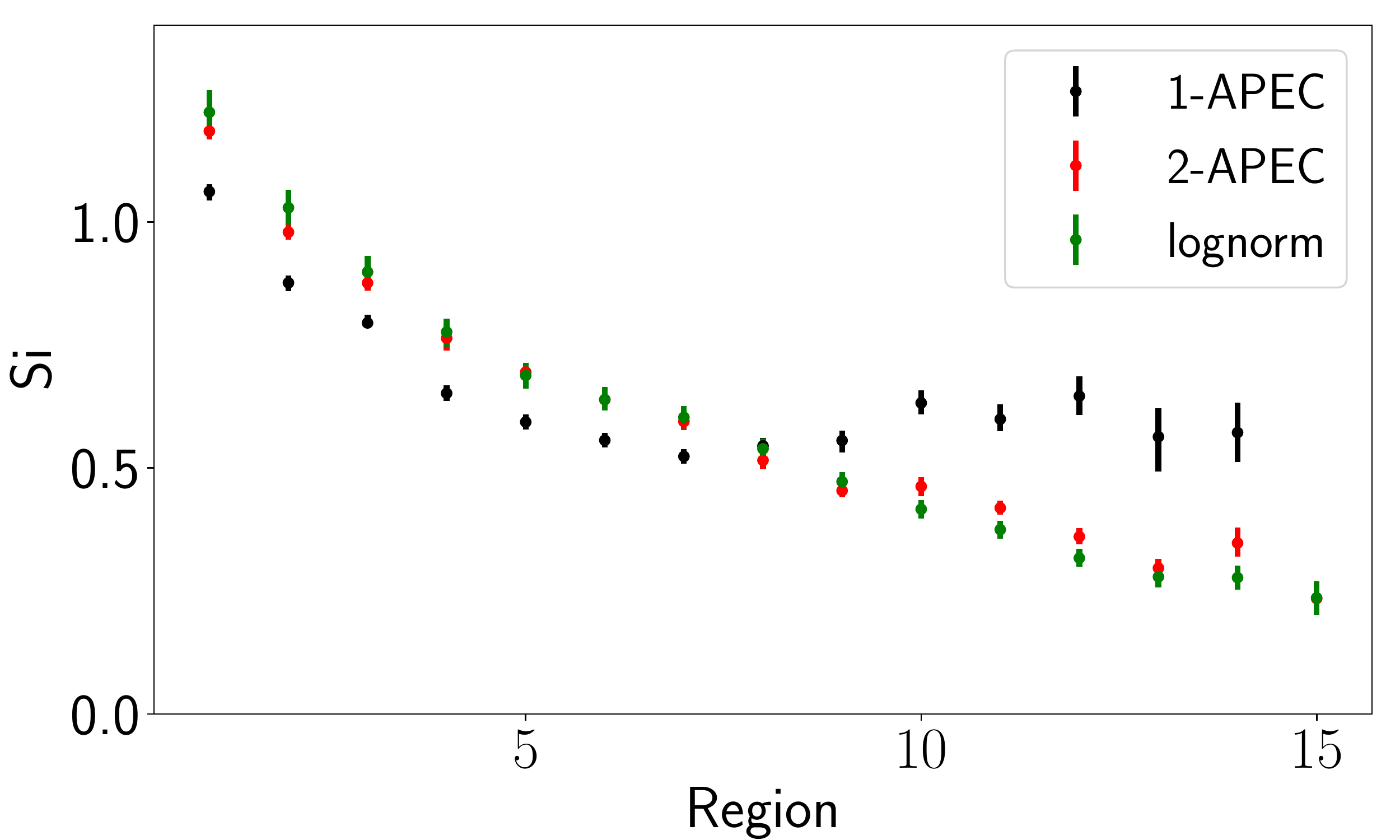} \\
\includegraphics[width=0.48\textwidth]{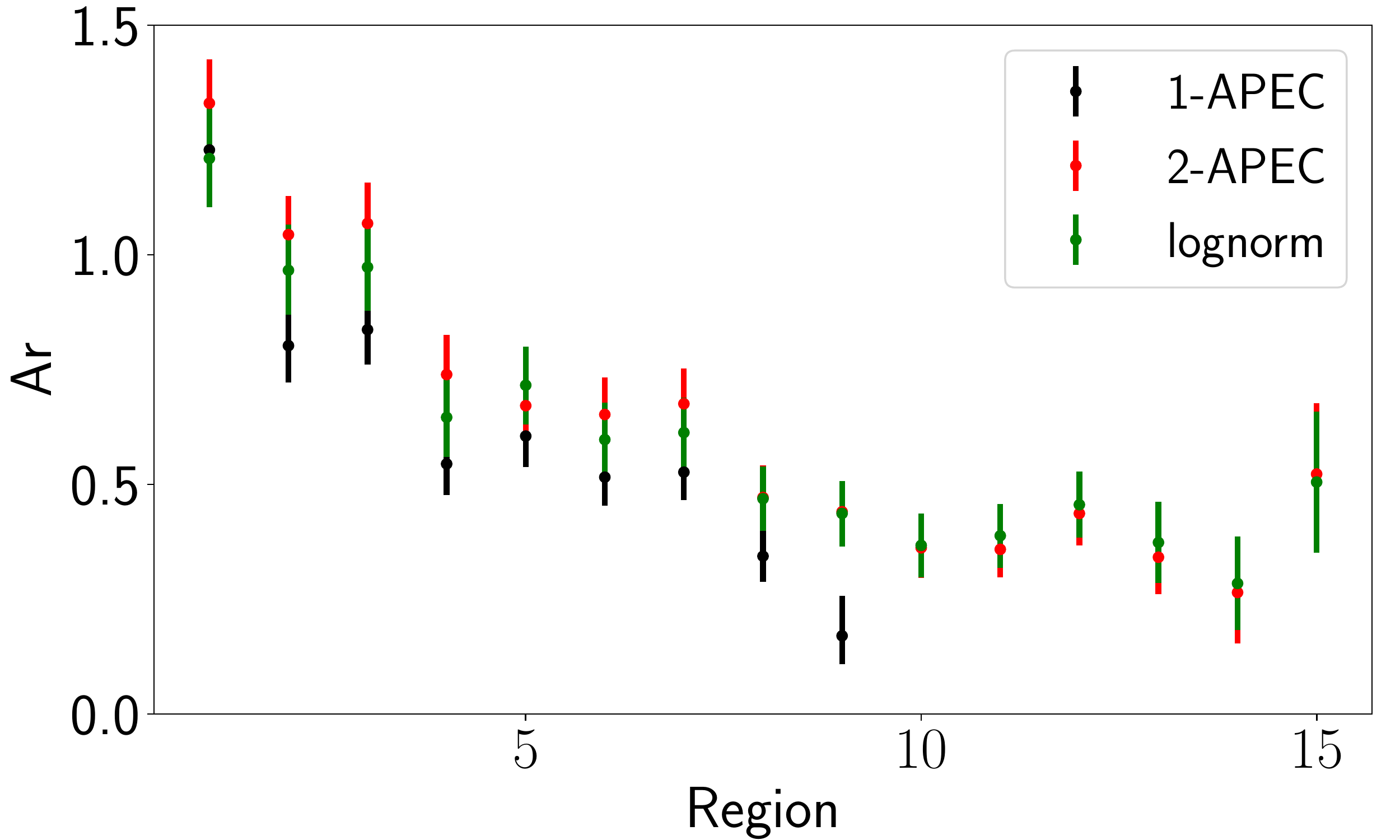} 
\includegraphics[width=0.48\textwidth]{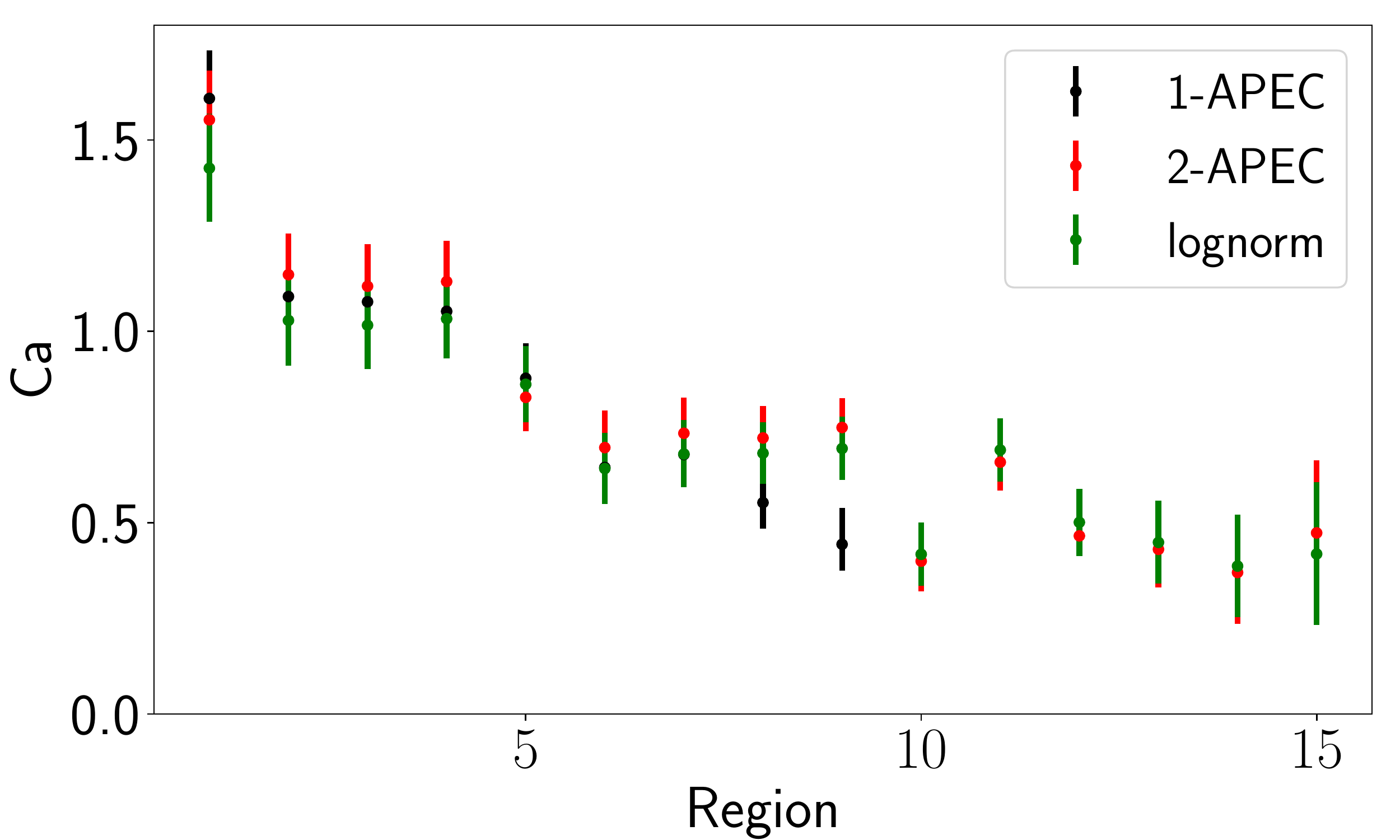} \\
\caption{S, Si, Ar and Ca best-fit abundances obtained with the 1-{\tt apec}, 2-{\tt apec} and {\tt lognorm} models.  } \label{metals_comparison} 
\end{figure*}

We investigated the thermodynamical substructure in the ICM with different temperature approximations. Apart from the {\tt lognorm} we also fit the spectra using a single (1-{\tt apec}) and a two (2-{\tt apec}) thermal components model (i.e. {\tt apec}). While 1-{\tt apec} and 2-{\tt apec} models are commonly fitted, it has been shown that a lognormal distribution has a more physical meaning \citep{vij22}. We assumed that the metal abundances were the same for both thermal components. Figure~\ref{cstat_comparison} shows the best-fit cash statistic obtained with each model. We found that for all regions the best-fit statistic corresponds to the {\tt lognorm} model while the single-temperature component performs worse.  Moreover, we found that a good agreement in most of elemental abundances between {\tt lognorm} and 2-{\tt apec} models. For example, Figure~\ref{metals_comparison} shows a comparison between the S, Si, Ar and Ca best-fit abundances obtained with the 1-{\tt apec} (black), 2-{\tt apec} (red) and {\tt lognorm} (green) models. We note that for larger regions the 1-{\tt apec} best-fit cannot recover the Ar and Ca abundances (i.e. only upper limits with < 0.1 values are found). We also fitted the spectra with these 3 models using $N({\rm H})$ as free parameter. We found that the abundances profiles are similar to those obtained with $N({\rm H})$ fixed, except for oxygen, whose emission signatures are located at the lower energies. In that sense, we are less confident about the O constraints. Finally, when considering a 2-{\tt lognorm} model, the best-fit does not provide a good constrain for both $\log(\sigma)$ values. Given these results we decide to perform the analysis of the Virgo cluster using the {\tt lognorm} model best-fit results.

\end{document}